\documentclass[aip,jcp,reprint,superscriptaddress,amsmath,amssymb,floatfix]{revtex4-2}
\usepackage{graphicx}
\newcommand{\RR}{\raggedright\let\\\tabularnewline}
\usepackage{bm}
\usepackage[dvipsnames]{xcolor}
\usepackage[colorlinks=true,linkcolor=blue,citecolor=blue,urlcolor=blue]{hyperref}
\newcommand{\K}{\ensuremath{\,\mathrm{K}}}
\newcommand{\mNm}{\ensuremath{\,\mathrm{mN\,m^{-1}}}}
\newcommand{\mNmK}{\ensuremath{\,\mathrm{mN\,m^{-1}\,K^{-1}}}}

\begin{document}

\title{Does the surface-tension anomaly of supercooled water reach the oil--water interface?}

\author{Fawaz Hrahsheh}
\email[Author to whom correspondence should be addressed: ]{fhrahsheh@hct.ac.ae}
\affiliation{Higher Colleges of Technology, ETS, MZWC, Abu Dhabi, 58855, United Arab Emirates}
\author{Moaz M. Altarawneh}
\affiliation{Department of Physics, American University of Sharjah, P.O. Box 26666, Sharjah, United Arab Emirates}
\author{Gerald Wilemski}
\email{wilemski@mst.edu}
\affiliation{Department of Physics, Missouri University of Science and Technology, Rolla, Missouri 65409, USA}

\date{\today}

\begin{abstract}
Below about 250~K, the surface tension of supercooled water rises anomalously. Whether this anomaly belongs to the liquid, and hence to all its interfaces, or to the free surface alone is unknown. This question determines how supercooled water is wetted, because an oil's spreading coefficient is a small difference of three large tensions. To address it, we couple the two-state equation of state of real water to square-gradient PC-SAFT theory. With two cross parameters fixed at room temperature, the theory reproduces the measured slopes of the heptane-- and octane--water tensions between 293 and 343~K within 5\% and predicts wetting temperatures of $271\pm14$~K (heptane) and $362\pm20$~K (octane). In the theory, only 55\% of the anomalous rise of the water tension reaches the oil--water interface. The remaining 45\% enters the spreading coefficient, so the lens contact angle nearly stops growing on cooling below 250~K. This fraction, nearly the same for both alkanes and at all temperatures, is set by how the gradient energy of the oil--water interface is shared between the two liquids, not by the anomaly. Other ways of imposing the anomaly give 32--78\%, always a partial transmission. Molecular dynamics simulations of TIP4P/2005 water agree with the theory on the structure of interfacial water, a disordered outer layer over a tetrahedral subsurface, independent of chain length, but they do not resolve the fraction. Measuring the octane--water tension near 240~K to 0.1~mN\,m$^{-1}$, or the lens angle below 240~K, would determine it for real water.
\end{abstract}

\maketitle

\section{Introduction}

Liquid water behaves as a fluctuating mixture of two local structures, one low-density and tetrahedrally ordered, and the other denser and disordered.\cite{Tanaka2000,Russo2014,Gallo2016,Anisimov2018} On cooling, their balance shifts toward the tetrahedral structure. This shift accounts for the density maximum and the growing response functions of supercooled water, which may end in a liquid--liquid critical point (LLCP).\cite{Poole1992,Holten2014,Debenedetti2020,Kim2020} Two-state equations of state built on this picture describe the bulk thermodynamics of supercooled water within experimental uncertainty.\cite{Holten2014,IAPWSG12} The surface tension of supercooled water is also anomalous. Early measurements down to $-27\,^\circ$C\cite{Floriano1990} were followed by the capillary-rise and counter-pressure measurements of Hrub\'y, Vin\v{s} and co-workers, which found no inflection down to $-26\,^\circ$C.\cite{Hruby2014,Vins2015} However, when extended to $-31.4\,^\circ$C, these data show that below about 250\K\ the surface tension rises faster than any extrapolation from the stable liquid predicts.\cite{Vins2020} Two-structure models of this rise were proposed early on.\cite{HrubyHolten2004,Feeney2003} More recently, Longo \textit{et al.}\cite{Longo2023} applied density gradient theory to a lattice model of two interconverting species and found that the liquid--vapor surface tension has either an inflection point or two extrema in its temperature dependence. In addition, simulations attribute the rise to the structure of the free surface itself, either through a second inflection point,\cite{Malek2019,Hrahsheh2024} the adsorption of high-density water,\cite{Gorfer2023} or competing hydrogen-bond orders.\cite{Yuan2026}

It is not known whether the anomaly belongs to water itself, and therefore to all its interfaces, or to the water/vapor interface alone. This question has a measurable consequence. An oil placed on water spreads when the (initial) spreading coefficient
\begin{equation}
S=\gamma_w-\gamma_a-\gamma_{aw}
\label{eq:S}
\end{equation}
is non-negative. Otherwise, it forms lenses whose contact angle is given by the Neumann triangle,
\begin{equation}
\cos\theta_c=\frac{\gamma_w^2-\gamma_a^2-\gamma_{aw}^2}{2\gamma_a\gamma_{aw}},
\label{eq:neumann}
\end{equation}
where $\gamma_w$, $\gamma_a$ and $\gamma_{aw}$ are the water/vapor, alkane--vapor and alkane--water tensions, respectively. For alkanes on water, $S$ is a small difference of large tensions, and therefore the temperature $T_w$ at which $S=0$ is unusually sensitive to molecular details.\cite{Ragil1996,Shahidzadeh1998,Pfohl1999,Bertrand2000} For example, in simulations of TIP4P/2005 water, one methylene unit moves $T_w$ from 286\K\ (heptane) to 413\K\ (octane), and $T_w$ shifts by 50--60\K\ per mN\,m$^{-1}$ of change in the tensions.\cite{Hrahsheh2023,companionMD} For the same reason, an anomaly of $\gamma_w$ that is not matched by $\gamma_{aw}$ must appear directly in $S$ and in the lens angle, whereas an anomaly that $\gamma_{aw}$ shares completely with $\gamma_w$, that is, one that raises $\gamma_{aw}$ by the same amount, leaves $S$ unchanged and the lens angle practically unchanged. Therefore, the lens angle of an alkane on supercooled water probes how much of the surface-tension anomaly is inherited by the oil--water interface. Although simulations of alkanes on model water already show unusual wetting trends at low temperatures,\cite{Neupane2021,companionMD} to our knowledge, no theory exists for the wetting of supercooled real water by alkanes.

In this work, we provide such a theory and compare it with the simulations of our companion study of TIP4P/2005 water with TraPPE-UA heptane and octane over 225--440\K.\cite{companionMD} The theory couples the two-state equation of state of real water\cite{Holten2014} to the square-gradient theory of interfaces,\cite{Cahn1958,RagilJCP1996} in which the perturbed-chain statistical associating fluid theory (PC-SAFT) equation of state\cite{Gross2001,Gross2002} supplies the Helmholtz energy of the inhomogeneous fluid. Functionals based on PC-SAFT have been applied before to liquid--liquid interfaces,\cite{Klink2015} where the accuracy for aqueous mixtures was limited by the description of the phase equilibria. The present theory uses the measured surface tension of supercooled water as an input, together with two cross parameters fixed by room-temperature data. We find that the theory reproduces the measured temperature dependence of $\gamma_{aw}$ without adjustment, predicts wetting temperatures of 271 and 362\K, and predicts that the growth of the lens angle on cooling nearly stops below about 250\K. We then investigate what this leveling-off requires, and we show that it is the signature of a \emph{partial} transmission of the anomaly. In the theory, 55\% of the anomalous rise of $\gamma_w$ reappears in $\gamma_{aw}$. We trace this fraction to the structure of square-gradient theory rather than to the anomaly itself, we show that the simulations do not resolve the transmitted fraction, and we identify the measurements that would fix this fraction for real water. Finally, theory and simulation agree on the two-state structure of interfacial water, which turns out to be independent of the alkane chain length and does not feed back on the wetting.

\section{Theory}

\subsection{Two-state equation of state for bulk water}

\begin{figure}[t]
\includegraphics[width=\columnwidth]{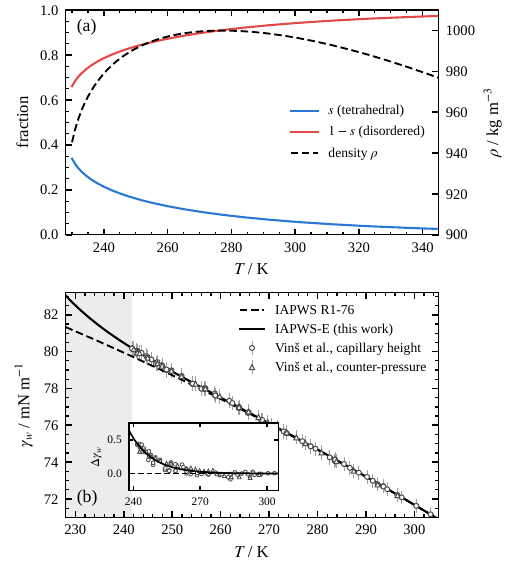}
\caption{Bulk two-state water and its surface tension. (a) Fractions of the tetrahedral ($s$) and disordered ($\rho$) structures and the density of real water at 0.1\,MPa from the two-state EOS.\cite{Holten2014} (b) Surface tension of supercooled water measured by Vin\v{s} \textit{et al.}\cite{Vins2020} using the capillary-height (circles) and counter-pressure (triangles) methods, compared with the IAPWS correlation\cite{IAPWSR1} (dashed) and with the extended correlation IAPWS-E, Eq.~(\ref{eq:iapwsE}), fitted in this work (solid). Inset: deviations from IAPWS. Below 241.7\K\ (shaded), IAPWS-E is an extrapolation.}
\label{fig:bulk}
\end{figure}

The bulk liquid is described by the two-state equation of state (EOS) of Holten, Sengers and Anisimov,\cite{Holten2014} which underlies Guideline G12-15 of the International Association for the Properties of Water and Steam (IAPWS).\cite{IAPWSG12} In this EOS, water is treated as an ideal-plus-regular mixture of a disordered, high-density structure ($\rho$ state) and a tetrahedral, low-density structure ($s$ state), with $s$-state fraction $s$. The equilibrium fraction minimizes the Gibbs energy at fixed $T$ and $P$,
\begin{equation}
L(T,P)+\ln\frac{s}{1-s}+\omega(P)\,(1-2s)=0,
\label{eq:twostate}
\end{equation}
where $L$ is the Gibbs-energy difference between the two structures in units of $RT$, and $\omega=2+\omega_0\hat p$, with $\hat p=(P-P_c)/(\rho_0RT_c)$, is their nonideality. All functions and parameters are given in Sec.~S1 of the supplementary material (SM). The EOS places an LLCP at $T_c=228.2$\K\ and $P_c=0$\,MPa. At 0.1\,MPa, it gives $s=0.34$ at 230\K\ and 0.026 at 345\K\ [Fig.~\ref{fig:bulk}(a)], and its density and compressibility reproduce the measurements down to the lowest temperatures at which they exist.\cite{Hare1987,Speedy1976}

\subsection{PC-SAFT and the water parameter sets}

The Helmholtz energy of the inhomogeneous fluid is taken from PC-SAFT,\cite{Gross2001,Gross2002} with the association term of Chapman \textit{et al.}\cite{Chapman1990} for water and the original non-associating parameters of Gross and Sadowski\cite{Gross2001} for heptane and octane (SM, Sec.~S2). Water carries four association sites (two donors and two acceptors, scheme 4C) with the parameters of Diamantonis and Economou\cite{Diamantonis2011} (Table~\ref{tab:inputs}). In addition, three other published sets are used as sensitivity tests. These are the non-polar four-site set of Rehner and Gross,\cite{Rehner2020} whose regression included surface-tension data, the four-site set of Rom\'an-Ram\'irez and Leeke,\cite{Roman2016} and the two-site (2B) set of Cameretti and Sadowski.\cite{Cameretti2008} In Sec.~\ref{sec:slope}, we test the four sets against the measured temperature dependence of $\gamma_{aw}$.

PC-SAFT alone does not reproduce the density anomaly of water. Therefore, the water segment diameter $\sigma_w$ is adjusted at every temperature so that the PC-SAFT liquid density equals that of real water, which is taken from IAPWS-95\cite{Wagner2002} above 273.16\K\ and from the two-state EOS at 0.1\,MPa below this temperature. This step imports the low-density expansion of supercooled water into the interfacial calculation exactly, and its consequences are isolated by a control calculation in Sec.~\ref{sec:transmission}. The single binary parameter $k_{ij}$ of the water--alkane dispersion energy is fitted at 298.15\K\ to the solubility of the alkane in water.\cite{Tsonopoulos1999} The resulting water solubility in the alkane is then a prediction, which agrees with experiment within a factor of two (Table~\ref{tab:inputs}). The implementation was verified against the open-source libraries teqp\cite{Bell2022} and FeOs\cite{Rehner2023} to all printed digits.

\begin{table*}
\caption{\label{tab:inputs}Inputs of the theory and how they are fixed (main parameter set). Everything else is predicted, namely $\gamma_{aw}$ at all temperatures other than 295.15\,K, and hence $S(T)$, $\theta_c(T)$ and $T_w$, the liquid--liquid compositions, and the interfacial profiles. No parameter was fitted to simulation data.}
\begin{ruledtabular}
\begin{tabular}{p{0.26\textwidth}p{0.36\textwidth}p{0.32\textwidth}}
\RR Input & \RR Fixed by & \RR Value\\
\hline
\RR Bulk water: $s$-state fraction and density & \RR two-state EOS\cite{Holten2014} at 0.1\,MPa (density from IAPWS-95\cite{Wagner2002} above 273.16\,K) & \RR $s=0.34$ (230\,K), 0.026 (345\,K); LLCP at 228.2\,K, 0\,MPa\\
\RR Water PC-SAFT parameters (4C-DE) & \RR Diamantonis and Economou\cite{Diamantonis2011} & \RR $m=2.1945$, $\varepsilon/k_B=141.66$\,K, $\varepsilon^{AB}/k_B=1804.2$\,K, $\kappa^{AB}=0.2039$\\
\RR Water segment diameter $\sigma_w(T)$ & \RR real-water liquid density at each $T$ & \RR 2.2288\,\AA\ (355\,K) to 2.3135\,\AA\ (230\,K)\\
\RR Alkane PC-SAFT parameters & \RR Gross and Sadowski\cite{Gross2001} & \RR SM, Sec.~S2\\
\RR Binary parameter $k_{ij}$ & \RR alkane solubility in water at 298.15\,K\cite{Tsonopoulos1999} & \RR 0.1650 (heptane), 0.1719 (octane); predicts $x_w=2.9\times10^{-4}$ in heptane at 295\,K (exp.\ $\approx5\times10^{-4}$)\\
\RR Influence parameters $c_w(T)$, $c_a(T)$ & \RR $\gamma_w$ (IAPWS-E, Table~\ref{tab:iapwsE}) and $\gamma_a$ (Mulero \textit{et al.}\cite{Mulero2012}) at each $T$, anchored at 295.15\,K\cite{Goebel1997} & \RR anchors $\gamma_w=72.52$, $\gamma_a=20.05$ (heptane) and 21.55\,mN\,m$^{-1}$ (octane)\\
\RR Cross parameter $\beta$ in $c_{aw}=(1-\beta)\sqrt{c_wc_a}$ & \RR $\gamma_{aw}$ at 295.15\,K\cite{Goebel1997} (51.9 and 52.5\,mN\,m$^{-1}$) & \RR 0.581 (heptane), 0.574 (octane)\\
\RR Anomaly of $\gamma_w$ & \RR fit of Eq.~(\ref{eq:iapwsE}) to the measurements of Vin\v{s} \textit{et al.}\cite{Vins2020} & \RR $c=0.099$\,K$^{-1}$, $T_e=233.4$\,K (Table~\ref{tab:iapwsE})\\
\RR Local two-state model & \RR not fitted (Sec.~\ref{sec:local}) & \RR coarse-graining radius 0.45\,nm, $n_c=4$\\
\end{tabular}
\end{ruledtabular}
\end{table*}

\subsection{Square-gradient theory of the interfaces}
\label{sec:sgt}

Planar interfaces are described by square-gradient theory,\cite{Cahn1958} with the grand-potential functional
\begin{equation}
\Omega[\mathbf n]/A=\int\left[\Delta\omega(\mathbf n)+\tfrac12\sum_{ij}c_{ij}\,n_i'n_j'\right]\mathrm dz,
\label{eq:sgt}
\end{equation}
where $\Delta\omega=f_0(\mathbf n)-\sum_i\mu_in_i+P$ is the excess grand-potential density of the homogeneous fluid at the composition $\mathbf n(z)$, and $c_{ij}$ are the influence parameters. For a pure fluid, the surface tension is $\gamma=\int\sqrt{2c\,\Delta\omega}\,\mathrm dn$ between the coexisting densities, and thus $\gamma\propto\sqrt c$. The influence parameters $c_w$ and $c_a$ of water and of each alkane are fitted at every temperature to the respective surface tensions, with $\gamma_w$ taken from the correlation of Sec.~\ref{sec:iapwsE} and $\gamma_a$ from the correlations of Mulero \textit{et al.},\cite{Mulero2012} both anchored to the values used by Goebel and Lunkenheimer at 295.15\K\ (Table~\ref{tab:inputs}).\cite{Goebel1997} For the water--alkane interface, the cross influence parameter is taken as $c_{aw}=(1-\beta)\sqrt{c_wc_a}$, and $\gamma_{aw}$ is the minimum of Eq.~(\ref{eq:sgt}) over all density paths between the two coexisting liquids, which is found numerically as described in the SM (Sec.~S4). The single cross parameter $\beta$ is fitted to the measured $\gamma_{aw}$ at 295.15\K.\cite{Goebel1997} At every other temperature, $\gamma_{aw}$ is a prediction.

\subsection{Surface tension of supercooled water}
\label{sec:iapwsE}

Following Rogers \textit{et al.},\cite{Rogers2016} we represent the anomaly of $\gamma_w$ by adding an exponential term to the IAPWS correlation,\cite{IAPWSR1}
\begin{equation}
\gamma_w(T)=B\tau^{\mu}(1+b\tau)+\gamma_s\,e^{-c(T-T_e)},\quad \tau=1-\frac{T}{T_c^{lv}},
\label{eq:iapwsE}
\end{equation}
where we keep the IAPWS values $B=235.8\mNm$, $b=-0.625$, $\mu=1.256$ and $T_c^{lv}=647.096\K$ (the liquid--vapor critical temperature) and set $\gamma_s=1\mNm$, since only $\gamma_s e^{cT_e}$ enters. (Here, $c$ and $\mu$ are the conventional symbols of the IAPWS form. They are unrelated to the influence parameters $c_{ij}$ and the chemical potentials $\mu_i$, just as the two-state nonideality $\omega$ is unrelated to the grand-potential density $\Delta\omega$.) A weighted fit to all 85 points of Vin\v{s} \textit{et al.}\ (241.7--303.3\K)\cite{Vins2020} gives the parameters of Table~\ref{tab:iapwsE}. Since the scatter of the data is much smaller than their stated uncertainties, which are dominated by systematic contributions, we quote the parameter uncertainties both ways and propagate both in Sec.~\ref{sec:robust}. The fit removes the systematic deviation of the data from IAPWS below 253\K\ [Fig.~\ref{fig:bulk}(b)]. Above 273\K, the two correlations agree within 0.02\mNm, whereas below 260\K\ the anomaly grows quickly, mainly in the slope (Table~\ref{tab:iapwsE}). We call this fit IAPWS-E, and we use it, shifted by $+0.085\mNm$ to the 295.15\K\ anchor of Sec.~\ref{sec:sgt}, in all calculations unless stated otherwise. Below 241.7\K, it is an extrapolation, and Sec.~\ref{sec:robust} examines how much depends on its form.

\begin{table}
\caption{\label{tab:iapwsE}The extended correlation IAPWS-E, Eq.~(\ref{eq:iapwsE}), fitted to the 85 measurements of Vin\v{s} \textit{et al.}\cite{Vins2020} (241.7--303.3\,K), compared with IAPWS.\cite{IAPWSR1} The parameter uncertainties are obtained from the scatter of the data and, in parentheses, from their stated uncertainties. $\Delta\gamma_w=\gamma_w^{\mathrm{E}}-\gamma_w^{\mathrm{IAPWS}}$. Values below 241.7\,K are extrapolated.}
\begin{ruledtabular}
\begin{tabular}{lccc}
\multicolumn{4}{l}{$c=0.099\pm0.006$ ($\pm0.03$)\,K$^{-1}$;\ \ $T_e=233.4\pm0.7$ ($\pm3.6$)\,K}\\
\multicolumn{4}{l}{correlation of $c$ and $T_e$: 0.94;\ \ $\chi^2_\nu=0.04$}\\
\multicolumn{4}{l}{rms deviation / mN\,m$^{-1}$: IAPWS 0.19, IAPWS-E 0.041}\\
\multicolumn{4}{l}{\quad below 253\,K: IAPWS 0.33, IAPWS-E 0.042}\\
\hline
 & & \multicolumn{2}{c}{$-\mathrm d\gamma_w/\mathrm dT$ / mN\,m$^{-1}$\,K$^{-1}$}\\
$T$ / K & $\Delta\gamma_w$ / mN\,m$^{-1}$ & IAPWS & IAPWS-E\\
\hline
260 & 0.07 & 0.132 & 0.139\\
250 & 0.20 & 0.126 & 0.145\\
240 & 0.52 & 0.119 & 0.171\\
230 & 1.40 & 0.113 & 0.251\\
\end{tabular}
\end{ruledtabular}
\end{table}

\subsection{Local two-state model of the interface}
\label{sec:local}

To resolve the interfaces into the two structures, we evaluate Eq.~(\ref{eq:twostate}) with the local water density $\bar\rho(z)$, coarse-grained over the first two coordination shells (a sphere of radius $R=0.45$\,nm). Two effects enter. First, a stretched local environment favors the open structure. This effect is described by mapping $\bar\rho$ onto a local pressure by linear response, which saturates at the stability limit of the EOS. Second, missing neighbors suppress the open structure. This effect is described by a coordination penalty $-n_c\ln(\bar\rho/\rho_l)$, where $\rho_l$ is the bulk liquid density and $n_c=4$, which is added to $L$. Neither parameter was fitted, and their effect is examined, together with the two-site water profiles, in Sec.~S11 of the SM. The model is a post-processing of the square-gradient density profiles. Since it does not feed back on the free energy, the wetting results of Secs.~\ref{sec:slope}--\ref{sec:robust} are independent of it.

\subsection{Simulation data}

The simulation results are taken from the companion study,\cite{companionMD} which reports the three tensions, $S(T)$ and $\theta_c(T)$ of TIP4P/2005 water with TraPPE-UA heptane (225--345\K) and octane (225--440\K), together with a two-state classification of the interfacial water molecules. We use its simulated tensions (Tables S1--S3 of its supplementary material) and its correlations (its Table II) as given, and the methods and uncertainties are described in the companion paper. In that paper, the simulated tensions were correlated with the same functional form as Eq.~(\ref{eq:iapwsE}). Where derivatives of these correlations are shown below, this is noted.

\subsection{What is imposed and what is predicted}

Table~\ref{tab:inputs} summarizes what is imposed. The temperature-by-temperature adjustment of $\sigma_w$, $c_w$ and $c_a$ removes the known deficiencies of PC-SAFT for pure water. However, with the cross parameters fixed, the temperature dependence of $\gamma_{aw}$ follows from the PC-SAFT mixture free energy and the combining rule alone. It is this dependence that can be tested by experiment.

\section{Results}

\subsection{The oil--water tension and its temperature dependence}
\label{sec:slope}

\begin{figure*}[t]
\includegraphics[width=\textwidth]{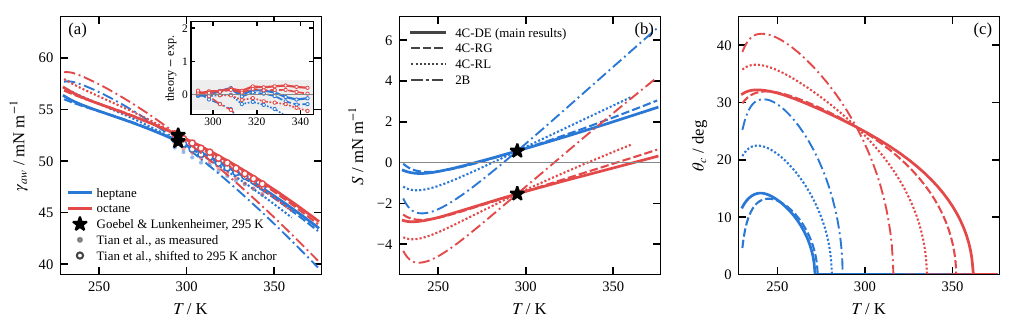}
\caption{Wetting of real water ($\gamma_w$ from IAPWS-E) with four PC-SAFT parameter sets for water (Table~\ref{tab:sets}): 4C-DE (solid; main results), 4C-RG (dashed), 4C-RL (dotted) and 2B (dash-dotted). (a) Alkane--water tension, with the data of Tian \textit{et al.}\cite{Tian2014} shifted to the 295.15\K\ anchor (open circles) and as measured (small filled circles). Stars are the anchor values of Goebel and Lunkenheimer.\cite{Goebel1997} Inset: theory minus the anchored data for the four sets (same line styles). The gray band is the expanded experimental uncertainty of $\pm0.45\mNm$. (b) Spreading coefficient. (c) Lens contact angle.}
\label{fig:models}
\end{figure*}

The least certain ingredient of any prediction of $S(T)$ is $\mathrm d\gamma_{aw}/\mathrm dT$, which is nearly as large as $\mathrm d\gamma_w/\mathrm dT$ and has the same sign. Tian \textit{et al.}\cite{Tian2014} measured $\gamma_{aw}$ for heptane and octane on water at eleven temperatures between 293 and 343\K\ with an expanded uncertainty of 0.45\mNm. Their absolute values lie 0.8--1.0\mNm\ below those of Goebel and Lunkenheimer,\cite{Goebel1997} which we use as anchors (the spread between published data sets\cite{Goebel1997,Zeppieri2001,Tian2014} is of this order). Therefore, the data were shifted by a constant to the anchor before comparison. Figure~\ref{fig:models} and Table~\ref{tab:sets} compare the four water parameter sets with the measurement. The main set predicts the measured slope within 5\% and reproduces the eleven individual values within 0.28\mNm. Its octane residuals grow slightly with temperature, which reflects a 4\% slope difference. The Rehner--Gross set passes this test equally well, whereas the Rom\'an-Ram\'irez--Leeke set is 11--20\% too steep and the two-site set is 46--58\% too steep. Two simpler estimates lie on either side of the measurement. The Fowkes--Good combination\cite{Girifalco1957,Fowkes1964} of the pure-component tensions is too steep, while the TIP4P/2005 simulations\cite{companionMD} are too shallow (Table~\ref{tab:sets}). The four sets differ mainly in the water dispersion energy, and hence in the fitted cross-dispersion energy $\varepsilon_{wa}=(1-k_{ij})\sqrt{\varepsilon_w\varepsilon_a}$. The two-site set, which has by far the strongest water--alkane attraction, lets $\gamma_{aw}$ follow $\gamma_w$ most closely and gives the steepest slope. Among the three four-site sets, which also differ in segment number and diameter, the slope does not follow $\varepsilon_{wa}$ alone. For all results below, we use the set of Diamantonis and Economou, which was selected before the Rehner--Gross set was tested. The latter set serves as an independent check of every conclusion (Table~\ref{tab:sets}).

To test whether this agreement transfers to other alkanes, we repeated the calculation with the main set for hexane and nonane, which Tian \textit{et al.}\cite{Tian2014} measured with the same method (SM, Sec.~S12). With the anchors of Goebel and Lunkenheimer\cite{Goebel1997} for these two alkanes, the predicted slope is $-0.092\mNmK$ for both, compared with the measured $-0.091$ and $-0.101\mNmK$, respectively. Thus, the theory reproduces the slopes of the four alkanes within 10\%. However, the predicted slope is nearly independent of chain length, whereas the measured slope steepens from hexane to nonane.

\begin{table*}
\caption{\label{tab:sets}Wetting of real water ($\gamma_w$ from IAPWS-E) with four published PC-SAFT parameter sets for water: four-site Diamantonis--Economou\cite{Diamantonis2011} (4C-DE), four-site Rehner--Gross\cite{Rehner2020} (4C-RG), four-site Rom\'an-Ram\'irez--Leeke\cite{Roman2016} (4C-RL) and two-site Cameretti--Sadowski\cite{Cameretti2008} (2B). $\varepsilon_{wa}$ is the water--heptane cross-dispersion energy that follows from the fitted $k_{ij}$. Slopes are given in mN\,m$^{-1}$\,K$^{-1}$, and the measured slopes over 293--343\,K are $-0.091$ (heptane) and $-0.097$ (octane).\cite{Tian2014} The rms is the root-mean-square deviation (mN\,m$^{-1}$) from the eleven measured values after anchoring. $\Delta\gamma_{aw}^{\mathrm{exp}}$ is the change of $\gamma_{aw}$ at 230\,K when the low-density expansion of water is suppressed (control A, Sec.~\ref{sec:transmission}), and n.c.\ means not computed. For comparison, the Fowkes--Good estimate\cite{Girifalco1957,Fowkes1964} gives slopes of $-0.116$ (heptane) and $-0.120$ (octane), and the TIP4P/2005 simulations\cite{companionMD} give $-0.062$ and $-0.064$\,mN\,m$^{-1}$\,K$^{-1}$ (285--345\,K).}
\begin{ruledtabular}
\begin{tabular}{l cccc}
 & 4C-DE & 4C-RG & 4C-RL & 2B\\
\hline
$\varepsilon_w/k_B$ (water) / K & 141.7 & 169.8 & 150.1 & 353.9\\
$\varepsilon_{wa}/k_B$ (heptane) / K & 153 & 177 & 159 & 286\\
$\mathrm d\gamma_{aw}/\mathrm dT$, 293--343\,K (heptane/octane) & $-0.093/{-0.093}$ & $-0.097/{-0.097}$ & $-0.108/{-0.108}$ & $-0.143/{-0.143}$\\
rms vs.\ Tian \textit{et al.}\ (heptane/octane) & 0.11/0.19 & 0.15/0.11 & 0.44/0.24 & 1.39/1.19\\
$x_w$ in heptane at 295\,K & $2.9\times10^{-4}$ & $4.7\times10^{-4}$ & $7.7\times10^{-4}$ & $5.4\times10^{-3}$\\
$T_w$ (heptane) / K & 271.4 & 272.9 & 281.0 & 287.3\\
$T_w$ (octane) / K & 361.7 & 352.1 & 335.3 & 316.1\\
$\theta_c$ maximum, heptane & $14.2^\circ$ (240\,K) & $13.2^\circ$ (245\,K) & $22.5^\circ$ (239\,K) & $30.6^\circ$ (242\,K)\\
$\theta_c$ maximum, octane & $32.2^\circ$ (238\,K) & $31.9^\circ$ (244\,K) & $36.6^\circ$ (238\,K) & $42.0^\circ$ (241\,K)\\
Transmitted fraction $f$, Eq.~(\ref{eq:f}) & 0.55 & 0.51 & n.c. & 0.45\\
$\Delta\gamma_{aw}^{\mathrm{exp}}$ at 230\,K / mN\,m$^{-1}$ & $+0.02$ & $+0.3$ & n.c. & $+1.7$\\
\end{tabular}
\end{ruledtabular}
\end{table*}

\subsection{Wetting temperatures}

\begin{figure*}[t]
\includegraphics[width=\textwidth]{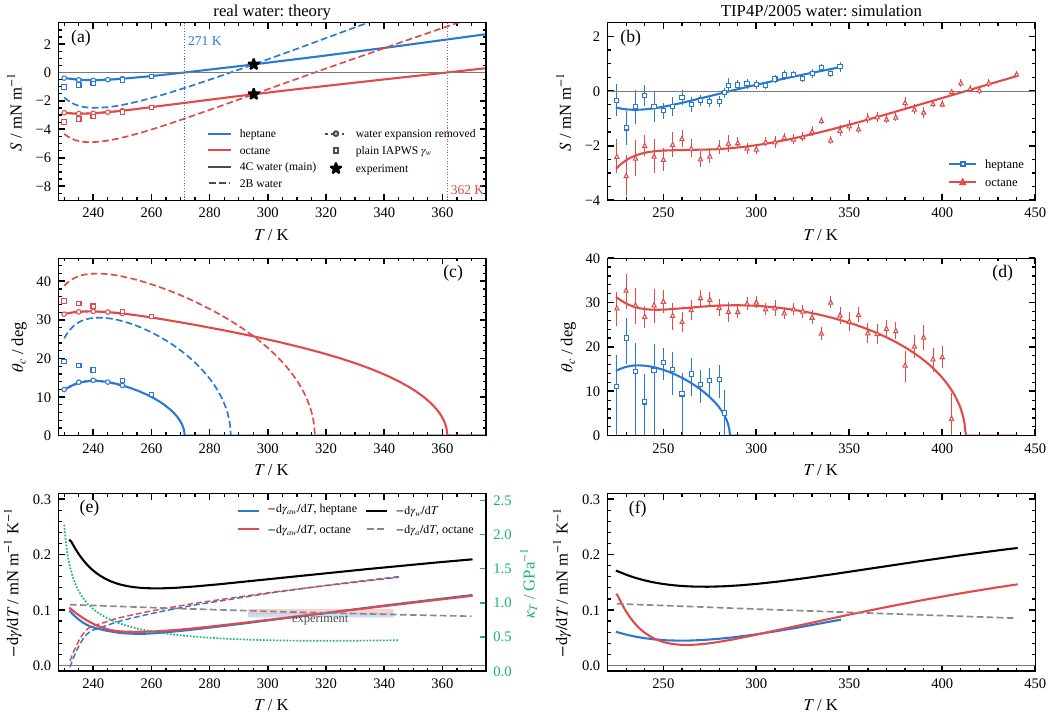}
\caption{Wetting of water by heptane (blue) and octane (red): theory for real water (left) and molecular dynamics of TIP4P/2005 water\cite{companionMD} (right). (a), (b) Spreading coefficient; (c), (d) lens contact angle; (e), (f) temperature derivatives of the tensions. Theory: solid lines, four-site water\cite{Diamantonis2011} with IAPWS-E; dashed, two-site water;\cite{Cameretti2008} stars, experimental values at 295.15\K;\cite{Goebel1997} dotted vertical lines, $T_w$; circles, control A (water segment diameter frozen at its 250\K\ value, so that below 250\K\ water contracts on cooling instead of expanding; in (c) they are indistinguishable from the lines); squares, control B (plain IAPWS $\gamma_w$). In (e), the colored solid lines are $-\mathrm d\gamma_{aw}/\mathrm dT$ for the four-site water and the thin colored dashed lines the same for the two-site water; $-\mathrm d\gamma_a/\mathrm dT$ is drawn for octane only (that of heptane lies within 0.01\mNmK\ of it); the shaded bands are the slopes of $\gamma_{aw}$ measured over 293--343\K\cite{Tian2014} and the green dotted line is the isothermal compressibility of bulk water (right axis), shown for reference. Simulation: symbols from the simulated tensions at each temperature, with standard errors for $S$ and 68\% Monte Carlo intervals for $\theta_c$;\cite{companionMD} lines from their correlations, which share the functional form of Eq.~(\ref{eq:iapwsE}); (f) uses the line styles of (e), with $-\mathrm d\gamma_w/\mathrm dT$ from the 225--440\K\ water correlation (the 225--345\K\ correlation of $\gamma_w$ differs from it by at most 0.11\mNm).}
\label{fig:wet}
\end{figure*}

Figure~\ref{fig:wet}(a) shows $S(T)$, and Table~\ref{tab:compare} collects the wetting results together with experiment and simulation. With the anchored tensions, $S=+0.57\mNm$ (heptane) and $-1.53\mNm$ (octane) at 295.15\K. Heptane, which spreads at room temperature ($S>0$), has $S=0$ at $T_w=271\K$, whereas octane forms lenses at room temperature and reaches $S=0$ at $T_w=362\K$. Two remarks are needed on the meaning of these temperatures. First, $S$ in Eq.~(\ref{eq:S}) is the initial spreading coefficient, which is formed with the tension of pure water. The equilibrium spreading coefficient of mutually saturated phases is never positive, and alkanes with $S>0$ spread to a film that coexists with lenses (pseudo-partial wetting) rather than to a macroscopic layer.\cite{RagilJCP1996,BonnRoss2001} As used here, $T_w$ is the temperature at which the initial $S$ changes sign, which is the quantity accessible to the room-temperature measurements\cite{Goebel1997,Tian2014} and to the simulations.\cite{companionMD} The first-order and critical wetting transitions of alkanes on water and brine, which involve the long-range interactions in the film,\cite{Ragil1996,Shahidzadeh1998,Bertrand2000,BonnRoss2001} are beyond the reach of a square-gradient theory of this kind. Second, because $S$ crosses zero with a slope of only $0.023\mNmK$, $T_w$ is very sensitive to the inputs (SM, Sec.~S10). It moves by about 43\K\ per mN\,m$^{-1}$ of error in $S$, which is close to the 50--60\K\ per mN\,m$^{-1}$ of the simulations.\cite{companionMD} We take $\pm0.3\mNm$ as the uncertainty of the absolute $\gamma_{aw}$ anchor (Tian \textit{et al.}\ quote 0.45\mNm, and the two data sets differ by 0.8--1.0\mNm\ before anchoring) and $\pm0.004\mNmK$, which is half the difference between the two measured alkane slopes, as the uncertainty of its slope. Table~\ref{tab:compare} gives the resulting uncertainty budget. For octane, a slope steeper by $0.004\mNmK$ moves the zero of $S$ above the highest computed temperature (375\K), and therefore the upper error of 20\K\ contains an estimated slope contribution of 15\K\ (SM, Sec.~S10). The two parameter sets that pass the slope test agree within these ranges (Table~\ref{tab:sets}), whereas the sets that fail it give an octane $T_w$ outside them.

\begin{table*}
\caption{\label{tab:compare}Wetting of water by heptane and octane: experiment, theory for real water (main parameter set) and TIP4P/2005 simulations.\cite{companionMD} Values are given as heptane/octane, with tensions in mN\,m$^{-1}$ and slopes in mN\,m$^{-1}$\,K$^{-1}$. The simulated $S$ at 295\,K and the slopes of $S$ are taken from the correlations of Ref.~\citenum{companionMD}, and the slopes of $\gamma_{aw}$ from its weighted linear fits.}
\begin{ruledtabular}
\begin{tabular}{p{0.29\textwidth}p{0.18\textwidth}p{0.23\textwidth}p{0.23\textwidth}}
 & \RR Experiment & \RR Theory (real water) & \RR Simulation (TIP4P/2005)\\
\hline
\RR $S$ at 295\,K & \RR $+0.57/{-1.53}$\cite{Goebel1997} & \RR $+0.57/{-1.53}$ (anchored) & \RR $+0.15/{-2.03}$\\
\RR $\mathrm d\gamma_{aw}/\mathrm dT$ (simulation: 285--345\,K; otherwise 293--343\,K) & \RR $-0.091/{-0.097}$\cite{Tian2014} & \RR $-0.093/{-0.093}$ & \RR $-0.062/{-0.064}$\\
\RR $\mathrm dS/\mathrm dT$ at $T_w$ & \RR -- & \RR $0.023/0.023$ & \RR $0.017/0.020$\\
\RR Shift of $T_w$ per mN\,m$^{-1}$ in $S$ / K & \RR -- & \RR $\approx43$ & \RR 50--60\\
\RR $T_w$ / K & \RR -- & \RR $271\pm14$ / $362\pm20$ & \RR $286\pm4$ / $413\pm3$\\
\RR \quad from $\pm0.3$\,mN\,m$^{-1}$ in the $\gamma_{aw}$ anchor & & \RR $\pm13$ / $\pm13$ & \\
\RR \quad from $\pm0.004$\,mN\,m$^{-1}$\,K$^{-1}$ in the slope of $\gamma_{aw}$ & & \RR $\pm$4--5 / $\pm15$ & \\
\RR $T_w^{\mathrm{octane}}-T_w^{\mathrm{heptane}}$ / K & \RR -- & \RR 90 & \RR 127\\
\RR $\theta_c$ on cooling, octane & \RR -- & \RR levels off below 250\,K at 31.5--32.2$^\circ$ & \RR levels off below 300\,K at $29.2^\circ$ (s.d.\ $1.8^\circ$)\\
\RR $\theta_c$ on cooling, heptane & \RR -- & \RR maximum $14.2^\circ$ at 240\,K & \RR maximum $\approx16^\circ$ near 237\,K\\
\RR Transmitted fraction $f$, Eq.~(\ref{eq:f}) & \RR -- & \RR 0.55 (0.32--0.78 under other assumptions, Table~\ref{tab:fvar}) & \RR not resolved (estimates 0.3--2)\\
\end{tabular}
\end{ruledtabular}
\end{table*}

The 90\K\ separation of the two wetting temperatures, compared with 127\K\ in the simulations of TIP4P/2005 water,\cite{companionMD} does not come from the difference between the two spreading coefficients at room temperature. This difference is about 2.1\mNm\ in experiment and theory and 2.2\mNm\ in the simulations, where 70\% of it comes from the larger surface tension of octane.\cite{companionMD} Instead, the separation is set by how fast $S$ grows on heating. This growth is slower in the simulations because the simulated $\gamma_{aw}$ varies more slowly with temperature than the measured one (Table~\ref{tab:compare}). Therefore, the same difference in $S$ is spread over a wider temperature interval. Expressed through these slopes, the differences between theory and simulation, 15\K\ for heptane and 51\K\ for octane, correspond to only about 0.3 and 1\mNm\ in $S$. The two-site water, whose $\gamma_{aw}$ varies too fast, compresses the separation to 29\K.

\subsection{The lens angle levels off on supercooling}
\label{sec:plateau}

On cooling below $T_w$, the lens angle first increases, as expected [Fig.~\ref{fig:wet}(c)]. For octane, it increases from $24.9^\circ$ at 300\K\ to $29.4^\circ$ at 270\K\ and $31.7^\circ$ at 250\K. Below 250\K, however, its growth nearly stops, and it gains only $0.5^\circ$ (octane) and $1.2^\circ$ (heptane) more before reaching shallow maxima. Between 250 and 230\K, the octane angle stays within $31.5$--$32.2^\circ$ and the heptane angle within $11.7$--$14.2^\circ$, with shallow maxima at 238 and 240\K, respectively. Figure~\ref{fig:wet}(e) shows the balance behind this behavior. On cooling, $\gamma_a$ grows at a nearly constant rate and $\gamma_w$ at an increasing rate because of its anomaly, whereas the growth of $\gamma_{aw}$ first \emph{slows}, from 0.078--0.079\mNmK\ at 295\K\ to 0.057--0.061\mNmK\ at 255\K. Below 250\K, the anomalous growth of $\gamma_w$ begins to outpace the combined growth of $\gamma_a$ and $\gamma_{aw}$. As a result, $S$ stops decreasing and the lens stops steepening.

The slowing of $\gamma_{aw}$ between 295 and 255\K\ is a regular feature of the PC-SAFT mixture free energy and is unrelated to the anomaly. Below 240\K, $\gamma_{aw}$ steepens again, to 0.110--0.113\mNmK\ at 230\K\ [Fig.~\ref{fig:wet}(e)]. It is in this steepening that the question posed in the Introduction is decided, and we examine it next.

\subsection{How much of the anomaly reaches the oil--water interface}
\label{sec:transmission}

Two ingredients of the calculation change below 250\K, namely the anomalous rise of $\gamma_w$ (IAPWS-E) and the low-density expansion of water (the two-state density). We removed each of them in turn (Fig.~\ref{fig:wet}; SM, Sec.~S7).

\emph{Control A: no low-density expansion.} The calculation below 250\K\ was repeated with $\sigma_w$ frozen at its 250\K\ value, so that PC-SAFT water contracts on cooling like a normal liquid, with everything else unchanged. With the main parameter set, the oil--water tension and the angles are practically unchanged down to 230\K\ (Table~\ref{tab:checks}). This means that the 5\% expansion of water between 250 and 230\K\ leaves the oil--water tension untouched. This result is not universal, however. With the Rehner--Gross and two-site sets, the same control changes $\gamma_{aw}$ noticeably, and with the two-site set the change is large enough to turn the contact-angle maximum into a monotonic increase. The effect grows monotonically with the cross-dispersion energy $\varepsilon_{wa}$ (Table~\ref{tab:sets}). In other words, the more of the water cohesion the oil--water interface feels, the more the expansion weakens it. Thus, whether the low-density expansion weakens the real oil--water interface is the one qualitative point below 250\K\ on which the parameter sets disagree.

\begin{table*}
\caption{\label{tab:checks}Control calculations (Sec.~\ref{sec:transmission}) and robustness checks (Sec.~\ref{sec:robust}) of the lens-angle results. The main calculation (four-site water, IAPWS-E) gives $\theta_c=14.2^\circ$ (heptane) and $32.2^\circ$ (octane) at 240\,K, and $11.7^\circ$ and $31.5^\circ$ at 230\,K. MC denotes Monte Carlo propagation (4000 draws) of the covariance of $c$ and $T_e$ with $\gamma_{aw}$ fixed.}
\begin{ruledtabular}
\begin{tabular}{p{0.27\textwidth}p{0.68\textwidth}}
\RR Calculation & \RR Effect\\
\hline
\multicolumn{2}{l}{\emph{Controls}}\\
\RR A: no low-density expansion ($\sigma_w$ frozen at 250\,K; 1003.8 instead of 945.3\,kg\,m$^{-3}$ at 230\,K) & \RR main set: $\gamma_{aw}$ changes by at most 0.02\,mN\,m$^{-1}$ and $\theta_c$ by at most $0.3^\circ$; Rehner--Gross set: $+0.3$\,mN\,m$^{-1}$ at 230\,K ($1.7^\circ$ octane, $7^\circ$ heptane); two-site set: $+1.7$\,mN\,m$^{-1}$, maximum of $\theta_c$ becomes a monotonic increase\\
\RR B: no surface-tension anomaly (plain IAPWS $\gamma_w$, two-state density kept) & \RR angles keep increasing on cooling, to $19.2^\circ$ (heptane) and $35.0^\circ$ (octane) at 230\,K; $\gamma_{aw}$ lower than with IAPWS-E by 0.11, 0.28 and 0.77\,mN\,m$^{-1}$ at 250, 240 and 230\,K, i.e., $f=0.55$\\
\multicolumn{2}{l}{\emph{Robustness}}\\
\RR (i) IAPWS-E uncertainty, scatter-based (MC) & \RR $\pm0.2^\circ$ at 250\,K; $\pm0.5^\circ$ (heptane) and $\pm0.3^\circ$ (octane) at 240\,K; 95\% intervals at 230\,K: 5--15$^\circ$ and 30--33$^\circ$; maximum of $\theta_c$ at 231--243\,K\\
\RR (i) IAPWS-E uncertainty, stated experimental uncertainties (MC) & \RR 240\,K: 0--15.5$^\circ$ (heptane), 29--33$^\circ$ (octane); maximum anywhere in 230--250\,K; octane levels off [$\theta_c(240\,\mathrm K)\le\theta_c(250\,\mathrm K)+2^\circ$] in $>99\%$ of draws; heptane: in 25\% the angle still rises at 240\,K, in 30\% the lens spreads ($S\ge0$) between 240 and 230\,K\\
\RR (ii) form of the continuation below 241.7\,K: quadratic, linear, anomaly frozen at the IAPWS slope & \RR down to 240\,K within $0.4^\circ$ (octane) and $1^\circ$ (heptane); octane at 230\,K: $32.3^\circ$, $33.7^\circ$, $36.5^\circ$ instead of $31.5^\circ$\\
\RR (iii) IAPWS-95 instead of the two-state EOS (962.4 instead of 945.3\,kg\,m$^{-3}$ at 230\,K) & \RR $\gamma_{aw}$ changes by $<0.01$\,mN\,m$^{-1}$, $\theta_c$ by $<0.1^\circ$\\
\RR (iv) path discretization, 120 to 200 nodes & \RR $\gamma_{aw}$: 0.006--0.017\,mN\,m$^{-1}$; $\theta_c$: $\le0.25^\circ$; $T_w$: $\le0.4$\,K\\
\end{tabular}
\end{ruledtabular}
\end{table*}

\emph{Control B: no surface-tension anomaly.} The calculation was repeated with the plain IAPWS correlation for $\gamma_w$ (two-state density retained) at 230--260\K. In this case, the angles keep increasing on cooling (squares in Fig.~\ref{fig:wet}; Table~\ref{tab:checks}). Therefore, the leveling-off requires the anomaly of $\gamma_w$. However, control B also shows how the anomaly reaches $\gamma_{aw}$. The water tension $\gamma_w$ enters the interfacial calculation only through the influence parameter $c_w$, which is 3.5\% larger with IAPWS-E than with IAPWS at 230\K. In turn, $\gamma_{aw}$ feels $c_w$ through the combining rule $c_{aw}=(1-\beta)\sqrt{c_wc_a}$ and through the water-density gradient on the water side of the interface. As a result, $\gamma_{aw}$ is higher with IAPWS-E than with IAPWS, by 0.77\mNm\ at 230\K\ (SM, Sec.~S7). Defining the transmitted fraction
\begin{equation}
f\equiv\frac{\gamma_{aw}^{\mathrm{IAPWS\text-E}}-\gamma_{aw}^{\mathrm{IAPWS}}}{\gamma_w^{\mathrm{IAPWS\text-E}}-\gamma_w^{\mathrm{IAPWS}}},
\label{eq:f}
\end{equation}
we find $f=0.55\pm0.01$ at every temperature between 230 and 260\K\ and for both alkanes with the main parameter set, and $f=0.45$ with the two-site set (245--290\K). Therefore, in this theory, about half of the anomalous rise of $\gamma_w$ reappears in $\gamma_{aw}$, and the steepening of $\gamma_{aw}$ below 240\K\ noted in Sec.~\ref{sec:plateau} is exactly this transmitted part. The other half enters $S$, and it is this part that stops the growth of the lens angle.

In the theory, the value of $f$ is fixed by the structure of the functional, not by the anomaly. Along a solution of Eq.~(\ref{eq:sgt}), the local and gradient terms are equal, and therefore $\gamma_{aw}=\int\mathbf n'^{\mathrm T}\mathbf C\,\mathbf n'\,\mathrm dz$, where $\mathbf C$ is the influence matrix. Scaling $c_w$ by $1+\epsilon$ scales $c_{aw}$ by $(1+\epsilon)^{1/2}$ and leaves $c_a$ unchanged. Because the profile is stationary, its first-order response can be evaluated at fixed profile, which gives $\delta\gamma_{aw}=\tfrac12\epsilon\int(c_wn_w'^2+c_{aw}n_w'n_a')\,\mathrm dz$, while for the free surface $\delta\gamma_w=\tfrac12\epsilon\gamma_w$. Hence,
\begin{equation}
f=\frac{\gamma_{aw}}{\gamma_w}\,\bar w,\qquad
\bar w=\frac{\int(c_wn_w'^2+c_{aw}n_w'n_a')\,\mathrm dz}{\int\mathbf n'^{\mathrm T}\mathbf C\,\mathbf n'\,\mathrm dz},
\label{eq:fw}
\end{equation}
where $\bar w$ is the share of the gradient energy of the oil--water interface carried by the water gradient (counting half the cross term). Evaluated on the optimal paths, $\bar w=0.79$ and $\gamma_{aw}/\gamma_w=0.69$ at 240--250\K, and Eq.~(\ref{eq:fw}) gives $f=0.546$ and 0.552, which agree within 0.001 with the values found from the two full calculations (SM, Sec.~S7). Since both factors are nearly independent of temperature, $f$ is constant. Neither factor involves the anomaly. The factor $\tfrac12$ from the square root cancels between $\delta\gamma_{aw}$ and $\delta\gamma_w$, and $f$ is set by the share of the oil--water gradient energy carried by water under the geometric-mean combining rule and by the room-temperature structure of the interface, not by the anomaly itself.

The value $f=0.55$ rests on two choices, namely that the whole anomaly is carried by the water influence parameter and that the cross influence parameter follows the geometric-mean rule. To assess how much $f$ depends on these choices, we repeated the calculation under different assumptions (Table~\ref{tab:fvar}; SM, Sec.~S13). First, within the same route, $f$ depends only weakly on the parameters. The Rehner--Gross set gives $f=0.51$, and the two-site set, with a different path and a different $\beta$, gives $\bar w=0.63$ and $f=0.45$. Moreover, when $\beta$ of the main set is varied from 0.40 to 0.75 without refitting, which changes $\gamma_{aw}$ at 240\K\ from 47 to 63\mNm, $f$ changes only from 0.51 to 0.58. Second, the way in which the cross parameter shares the anomaly matters more. If $c_{aw}$ scales as $c_w^{\alpha}$, the share $\bar w$ in Eq.~(\ref{eq:fw}) is replaced by $\bar w_\alpha=\int(c_wn_w'^2+2\alpha c_{aw}n_w'n_a')\,\mathrm dz/\int\mathbf n'^{\mathrm T}\mathbf C\,\mathbf n'\,\mathrm dz$, where $\alpha=1/2$ is the geometric-mean rule. Since the two densities vary in opposite directions through the interface, the cross term is negative and $f$ decreases with $\alpha$, from 0.77--0.78 for $\alpha=0$, where the cross parameter does not share the anomaly, to 0.32--0.33 for $\alpha=1$. Third, we kept $c_w$ at its regular value and carried the anomaly in the bulk free energy of water instead. To do so, we increased either the association energy or the dispersion energy of water at each temperature, with $\sigma_w$ readjusted to the density, until the water tension equals IAPWS-E. These two routes give $f=0.74$--0.76 and 0.71--0.72, respectively. Therefore, the anomaly is only partly transmitted in every variant, with $f$ between 0.32 and 0.78, whereas the value 0.55 belongs to the standard square-gradient route. It is worth mentioning here that the temperature dependence of the combining rule has never been tested below 293\K\ for water--alkane interfaces. Above this temperature, the measured slope of $\gamma_{aw}$ (Sec.~\ref{sec:slope}) is reproduced, which means that whatever anomaly-independent temperature dependence the rule imparts is correct there. Below 250\K, however, the value of $f$ in the theory should be read as a property of the square-gradient description, not as a prediction about water.

\begin{table}
\caption{\label{tab:fvar}Transmitted fraction $f$, Eq.~(\ref{eq:f}), under different assumptions (SM, Sec.~S13). Values are given as heptane/octane at 240\,K unless noted. In the first two groups, the anomaly is carried by $c_w$ and $c_{aw}\propto c_w^{\alpha}$, and $f$ is evaluated on the optimal path with Eq.~(\ref{eq:fw}) and $\bar w_\alpha$. In the third group, $c_w$ keeps its regular value and one energy parameter of water is increased until $\gamma_w$ equals IAPWS-E, and $f$ follows from the recomputed $\gamma_{aw}$.}
\begin{ruledtabular}
\begin{tabular}{lc}
Assumption & $f$\\
\hline
\multicolumn{2}{l}{\emph{Geometric-mean rule ($\alpha=1/2$)}}\\
Main set (4C-DE) & 0.55/0.55\\
Rehner--Gross set (4C-RG) & 0.51/0.51\\
Two-site set (2B), heptane at 245\,K & 0.45\\
Main set, $\beta$ varied from 0.40 to 0.75 & 0.51--0.58\\
\multicolumn{2}{l}{\emph{Other sharing by the cross parameter (main set)}}\\
$c_{aw}$ does not share the anomaly ($\alpha=0$) & 0.77/0.78\\
$c_{aw}$ shares it fully ($\alpha=1$) & 0.33/0.32\\
\multicolumn{2}{l}{\emph{Anomaly in the bulk free energy of water (main set)}}\\
Association energy $\varepsilon^{AB}$ & 0.75/0.74\\
Dispersion energy $\varepsilon$ & 0.72/0.71\\
\end{tabular}
\end{ruledtabular}
\end{table}

\begin{figure*}[t]
\includegraphics[width=\textwidth]{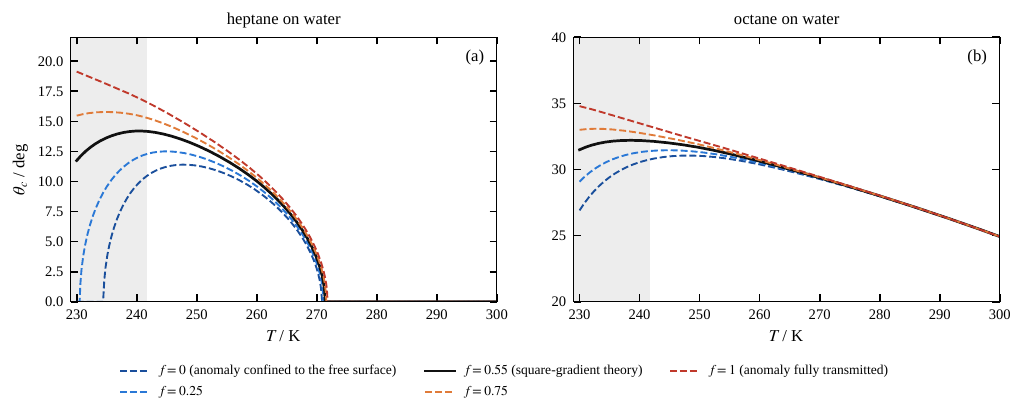}
\caption{The lens contact angle as a probe of the transmitted fraction $f$, Eq.~(\ref{eq:f}), for (a) heptane and (b) octane on water. Here, $\gamma_{aw}^{(f)}(T)=\gamma_{aw}^{\mathrm{IAPWS}}(T)+f\,[\gamma_w^{\mathrm{IAPWS\text-E}}(T)-\gamma_w^{\mathrm{IAPWS}}(T)]$, with $\gamma_w$ from IAPWS-E and $\gamma_a$ unchanged, and $\gamma_{aw}^{\mathrm{IAPWS}}$ is the tension of control B. The value $f=0.55$ (solid) is the square-gradient result, and $f=0.25$ and 0.75 (dashed) are intermediate values. The value $f=0$ confines the anomaly to the free surface, whereas $f=1$ transmits it fully. Below 241.7\K\ (shaded), $\gamma_w$ is extrapolated.}
\label{fig:probe}
\end{figure*}

\begin{table}
\caption{\label{tab:probe}Lens contact angle (degrees) of heptane and octane on water for different transmitted fractions $f$ of the surface-tension anomaly, Eq.~(\ref{eq:f}) and Fig.~\ref{fig:probe}. The value $f=0.55$ corresponds to the square-gradient theory with the main parameter set. The value $f=1$ gives the same $S$ as control B (no anomaly in $\gamma_w$), and its angles coincide with those of control B to within $0.2^\circ$. Below 241.7\,K, the water surface tension is extrapolated.}
\begin{ruledtabular}
\begin{tabular}{l ccccc}
 & \multicolumn{5}{c}{$T$ / K}\\
 & 230 & 235 & 240 & 250 & 260\\
\hline
heptane, $f=0$ & 0.0 & 3.5 & 9.7 & 11.3 & 9.2\\
heptane, $f=0.55$ & 11.7 & 13.6 & 14.2 & 13.0 & 10.0\\
heptane, $f=1$ & 19.1 & 18.1 & 17.0 & 14.2 & 10.6\\
octane, $f=0$ & 26.9 & 29.4 & 30.6 & 31.0 & 30.4\\
octane, $f=0.55$ & 31.5 & 32.1 & 32.2 & 31.7 & 30.6\\
octane, $f=1$ & 34.8 & 34.2 & 33.5 & 32.2 & 30.8\\
\end{tabular}
\end{ruledtabular}
\end{table}

The lens angle then measures $f$ itself. Figure~\ref{fig:probe} and Table~\ref{tab:probe} show $\theta_c(T)$ for a family of oil--water tensions that share the no-anomaly baseline of control B and transmit the anomaly by a fraction $f$. For $f=1$, the anomaly cancels in $S$, and the angle keeps rising on cooling, as in control B. For $f=0$, the full anomaly enters $S$. In this case, the heptane lens spreads below 234\K, and the octane angle turns down by $4^\circ$ between 240 and 230\K. For intermediate $f$, the angle levels off, with a maximum whose depth grows as $f$ decreases. Even for $f=0.75$, which is close to the largest value of Table~\ref{tab:fvar}, the octane angle gains only $0.9^\circ$ between 250 and 240\K\ and then stays at $33.0^\circ$ down to 230\K\ (Fig.~\ref{fig:probe}). The spread between $f=0$ and $f=1$ is $2.9^\circ$ (octane) and $7^\circ$ (heptane) at 240\K, which is just below the lowest temperature at which $\gamma_w$ has been measured (241.7\K), and it is $8^\circ$ and $19^\circ$ at 230\K, where $\gamma_w$ is extrapolated. Thus, the lens angle is a weak but genuine probe of $f$ within the measured range, and it is a sensitive one if the anomaly continues below 241.7\K\ as IAPWS-E assumes. A direct measurement of $\gamma_{aw}(T)$ below 273\K\ provides a sharper test. For $f=0.55$, the octane--water tension at 240\K\ lies 0.28\mNm\ above the regular continuation of its stable-liquid behavior, and 0.77\mNm\ above it at 230\K. Resolving these differences requires a relative precision of about 0.1\mNm, which is better than the 0.45\mNm\ expanded uncertainty of existing temperature-dependent data.\cite{Tian2014}

\subsection{Robustness}
\label{sec:robust}

Four further checks bound the result (Table~\ref{tab:checks}; SM, Secs.~S7, S8 and S10). (i) \emph{Uncertainty of the IAPWS-E fit.} With the scatter-based uncertainties, the angles down to 240\K\ are determined to better than a degree. When the stated experimental uncertainties are taken at face value, the octane angle still levels off in practically all draws, whereas for heptane, whose $S$ is within about 0.5\mNm\ of zero, the anomaly may grow either too slowly to stop the angle by 240\K\ or fast enough to spread the lens. (ii) \emph{Form of the extrapolation below 241.7\K.} The angles down to 240\K\ hardly depend on it. However, whether the angle turns down or merely stops growing below 240\K\ does depend on it, and only measurements below 241.7\K\ can settle this point. (iii) \emph{Equation of state and LLCP.} The two-state EOS places the LLCP at 0\,MPa, while recent estimates favor positive pressures.\cite{Debenedetti2020,Kim2020} The density of supercooled water at ambient pressure has been measured down to 239\K\cite{Hare1987} and its compressibility down to 247\K.\cite{Speedy1976} Therefore, any EOS that reproduces these data gives the same density anomaly there. Indeed, the two-state EOS and IAPWS-95,\cite{Wagner2002} which has no critical point, agree within 0.2\,kg\,m$^{-3}$ down to 235\K, and exchanging them changes nothing of consequence, as implied by control A. (iv) \emph{Numerical convergence.} The numerical convergence of $\gamma_{aw}$ is likewise immaterial. In summary, for the square-gradient value $f\approx0.5$, the leveling-off between 250 and 241.7\K\ follows from the measured $\gamma_w$ and density, to within $\pm2^\circ$ for octane even when the stated uncertainties of $\gamma_w$ are taken at face value. Its continuation below 241.7\K, however, is fixed by neither.

\subsection{Comparison with simulation}
\label{sec:md}

The TIP4P/2005 simulations\cite{companionMD} make no mean-field approximation and contain no combining rule, and therefore they test the picture from a different side [Figs.~\ref{fig:wet}(b), \ref{fig:wet}(d), \ref{fig:wet}(f); Table~\ref{tab:compare}]. Octane forms lenses on the model water up to $T_w\approx413\K$, but below 300\K\ its contact angle stops increasing and levels off, with no resolved temperature dependence down to 225\K. By contrast, heptane, whose $T_w$ is $\approx286\K$, shows no plateau above 235\K, and its angle rises from zero at $T_w$ to a shallow maximum near 237\K. Thus, the leveling-off of the octane angle is an observable common to theory and simulation, but the two reach it by different routes. In the theory, $S$ stops decreasing because 45\% of the water anomaly enters it. In the simulations, on the other hand, the alkane side of $S$ varies on cooling as fast as $\gamma_w$ or faster. Between 225 and 285\K, the weighted linear slopes are $-0.145\mNmK$ for $\gamma_w$ and $-0.156$ (heptane) and $-0.150\mNmK$ (octane) for $\gamma_a+\gamma_{aw}$,\cite{companionMD} and the correlated $\gamma_{aw}$ of both systems have an inflection near 260--263\K, close to that of the simulated water tension (270--272\K). Thus, the simulated octane plateau results from a balance of slopes. The octane $S$ is nearly flat because $\gamma_w$ and the alkane side change at nearly the same rate, whereas the heptane $S$ keeps falling because its alkane side changes faster, mainly through $\gamma_a$.\cite{companionMD} This balance does not by itself show how much of an anomaly is transmitted. To determine this, the anomalous parts of the simulated tensions would have to be separated from their regular parts, which cannot be done with any precision. Departures from a linear baseline fitted above 285\K\ are comparable with the statistical errors of the tensions at 225--250\K\ and are mostly negative, which reflects the regular curvature of the tensions rather than an anomalous rise. In addition, the ratio of the $\gamma_{aw}$ and $\gamma_w$ departures ranges from 0.3 to 1.6 depending on the alkane, the baseline window and the weighting of the fit, and derivatives of the correlations constrain it even less (SM, Sec.~S9). The simulations therefore do not resolve $f$, and the estimates range from 0.3 to about 2.

Thus, the simulations do not confirm the partial transmission found in the theory, but they do not exclude it either. They are compatible with a molecular model, with the same force-field physics at both interfaces, that transmits its surface anomaly fully. Whether real water behaves like the theory ($f\approx0.5$), like the model ($f\approx1$) or like neither is precisely what the measurements proposed above would decide. The derivative plot of Fig.~\ref{fig:wet}(f) should be read with this in mind. It is built from correlations that share the exponential form of Eq.~(\ref{eq:iapwsE}), and thus the fall of $-\mathrm d\gamma_{aw}/\mathrm dT$ from 0.07\mNmK\ at 325\K\ to 0.04 near 260\K\ and its rise again on further cooling are features of the fitted form constrained by the data, not model-free derivatives.

\begin{figure*}[t]
\includegraphics[width=\textwidth]{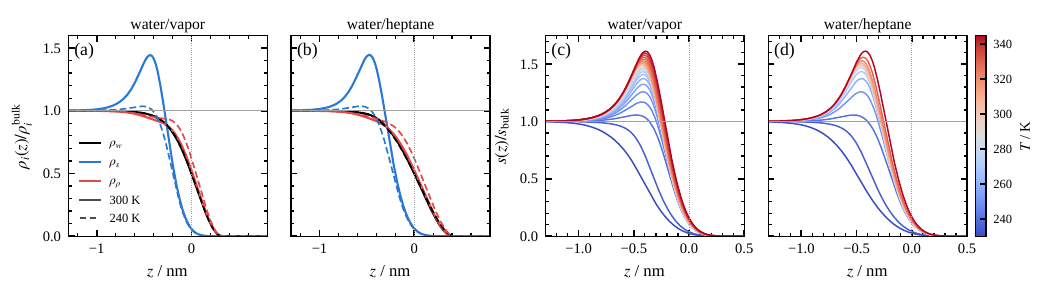}
\caption{Interfacial structure of real water (four-site water, IAPWS-E). (a), (b) Total water density $\rho_w$, $s$-state density $\rho_s=s\rho_w$ and $\rho$-state density $\rho_\rho=(1-s)\rho_w$, each relative to its bulk value, at 300\K\ (solid) and 240\K\ (dashed), next to vapor and next to heptane. (c), (d) Local $s$-state fraction relative to the bulk from 230 (blue) to 345\K\ (red). Liquid water is at $z<0$, and $z=0$ is the 50\% point of $\rho_w$. Profiles next to octane coincide with those next to heptane.}
\label{fig:prof}
\end{figure*}

\subsection{Two-state structure of interfacial water}
\label{sec:structure}

\begin{figure*}[t]
\includegraphics[width=\textwidth]{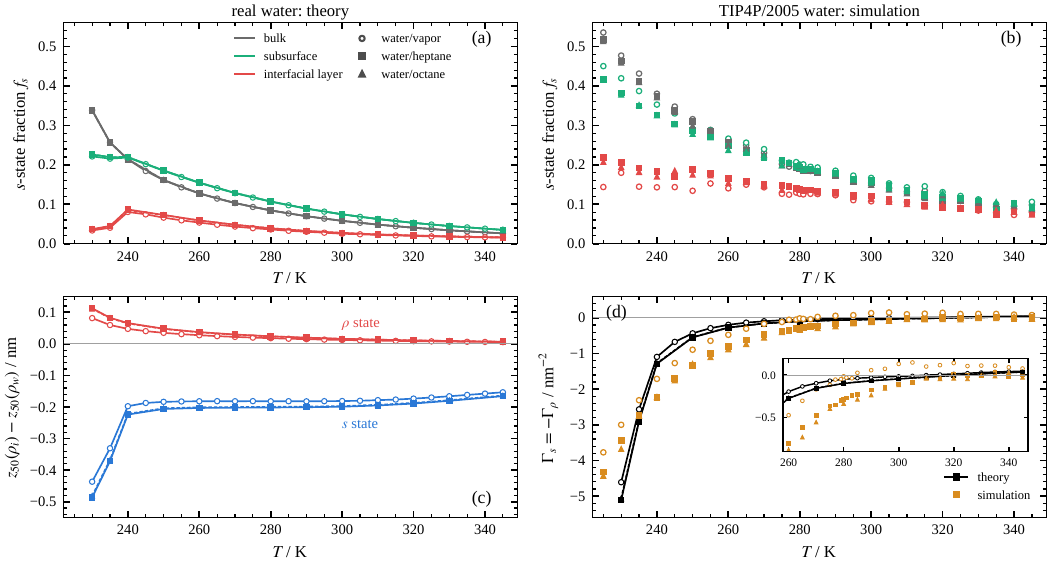}
\caption{Two-state balance of interfacial water against temperature next to vapor (open circles), heptane (squares) and octane (triangles; dashed in the theory). (a) Theory for real water: $s$-state fraction $f_s$ in the bulk, in the 0.5\,nm subsurface layer inside the 90\% point of $\rho_w$ and in the interfacial layer between the 90\% and 10\% points. (b) TIP4P/2005 simulation:\cite{companionMD} the corresponding fractions, with the subsurface layer taken between 1\,nm from the slab center and the 90\% point and the bulk fraction from a geometric classification of molecules. (c) Theory: position of the 50\% point of the $s$-state (blue) and $\rho$-state (red) profiles relative to that of $\rho_w$. (d) Relative Gibbs adsorption of $s$-state water from theory (black, lines) and simulation (orange, symbols only). The inset magnifies the sign change.}
\label{fig:sum}
\end{figure*}

Figure~\ref{fig:prof} resolves the interfaces into the two structures using the local model of Sec.~\ref{sec:local}. We describe $T\geq240\K$, where the trends are smooth, and Table~\ref{tab:struct} gives the numbers. The $\rho$-state profile follows the total density, with its 50\% point slightly \emph{outside} that of the total water. The $s$-state profile is sharper, and its 50\% point lies 0.15--0.22\,nm \emph{inside} the water [Fig.~\ref{fig:sum}(c)]. This means that the outermost layer is disordered water. In the interfacial layer between the 90\% and 10\% points of $\rho_w$, the $s$-state density is reduced to between a fifth and a third of its bulk value, whereas the $\rho$-state density is reduced only to about 0.6. Beneath the surface, on the other hand, the $s$ state is enriched. From 240\K\ on, a subsurface maximum of $\rho_s$ appears about 0.5\,nm below the 50\% point and grows on heating, to about 1.5 times the bulk value at 345\K\ (Table~\ref{tab:struct}; the local fraction $s$ itself peaks at 1.6 times its bulk value, Figs.~\ref{fig:prof}(c) and \ref{fig:prof}(d)). The $\rho$ state shows no such maximum, and in the subsurface layer it is slightly depleted. Both features follow from Eq.~(\ref{eq:twostate}). Beneath the surface, the coarse-grained density is slightly reduced but the coordination is nearly complete, so that the local stretch favors the open structure. Farther out, missing neighbors forbid tetrahedral order.

\begin{table*}
\caption{\label{tab:struct}Two-state structure of interfacial water for $T\ge240$\,K: theory for real water (four-site water, IAPWS-E) and TIP4P/2005 simulations (our analysis of the trajectories of Ref.~\citenum{companionMD}; SM, Sec.~S9). Positions are measured from the 50\% point of the total water density and are negative inside the liquid. The definitions of the regions differ in detail between theory and simulation (Fig.~\ref{fig:sum} and SM).}
\begin{ruledtabular}
\begin{tabular}{p{0.30\textwidth}p{0.14\textwidth}p{0.16\textwidth}p{0.14\textwidth}p{0.16\textwidth}}
 & \multicolumn{2}{c}{Theory} & \multicolumn{2}{c}{Simulation}\\
 & \RR next to vapor & \RR next to alkanes & \RR next to vapor & \RR next to alkanes\\
\hline
\RR 50\% point of $\rho_s$ / nm & \RR $-0.15$ to $-0.20$ & \RR $-0.17$ to $-0.22$ & \RR $-0.04$ to $-0.09$ & \RR $-0.03$ to $-0.09$\\
\RR 50\% point of $\rho_\rho$ / nm & \multicolumn{2}{c}{up to $+0.07$} & \RR $+0.01$ to $+0.05$ & \RR $+0.004$ to $+0.05$\\
\RR 10--90\% width of $\rho_s$ (of $\rho_w$) / nm & \RR 0.22--0.29 (0.48) & & & \\
\RR $\rho_s$ ($\rho_\rho$) in the interfacial layer, relative to bulk & \multicolumn{2}{c}{0.21--0.33 (0.54--0.64)} & & \\
\RR Subsurface maximum of $\rho_s$: depth / nm & \RR 0.40--0.52 & \RR 0.44--0.56 & & \\
\RR Subsurface maximum of $\rho_s$ / bulk, 300\,K (simulation: 295--300\,K) & \RR 1.44 & \RR 1.45 & \RR 1.17--1.20 & \RR 1.11--1.13\\
\RR $\Gamma_s=-\Gamma_\rho$ at 240\,K / nm$^{-2}$ & \RR $-1.10$ & \RR $-1.29$ & \RR $-1.71$ & \RR $-2.22$ to $-2.23$\\
\RR $\Gamma_s$ changes sign at / K & \RR 309 & \RR 321 & \RR $\approx285$ & \RR 315--321\\
\RR Heptane vs octane, largest difference in $f_s$ ($\Gamma_s$ / nm$^{-2}$) & & \RR 0.002 & & \RR 0.02 (0.23)\\
\end{tabular}
\end{ruledtabular}
\end{table*}

The two effects compete. With the dividing surface fixed by $\rho_w$, the relative adsorption of tetrahedral water is $\Gamma_s=\int[\rho_s(z)-s_\mathrm{bulk}\rho_w(z)]\,\mathrm dz=-\Gamma_\rho$ [Fig.~\ref{fig:sum}(d)]. At low temperature, the depletion of the outer layer dominates and $\Gamma_s<0$. On heating, the subsurface enrichment catches up, and $\Gamma_s$ changes sign at about 310--320\K\ (Table~\ref{tab:struct}). Above these temperatures, the interface holds a net excess of tetrahedral water, although its outermost layer remains depleted.

The simulations show the same trends [Figs.~\ref{fig:sum}(b), \ref{fig:sum}(d); Table~\ref{tab:struct}], with definitions that differ in detail (SM, Sec.~S9). The outermost layer of TIP4P/2005 water is depleted of the tetrahedral state at every temperature, and the alkane makes it more tetrahedral than vapor at low temperature. In addition, a subsurface enrichment develops on heating, which is weaker next to the alkane, the $s$-state front lies inside and the $\rho$-state front outside the total-density front (by a quarter to a third of the displacements in the theory), and the adsorption $\Gamma_s$ follows the theory in sign and temperature trend. Next to the alkanes, its magnitude exceeds that of the theory by an amount that grows on cooling from $0.2\,\mathrm{nm}^{-2}$ at 280\K\ to $0.95\,\mathrm{nm}^{-2}$ at 240\K. The simulated bulk fractions are larger (0.51--0.54 at 225\K) because the geometric classification counts more molecules as tetrahedral than the thermodynamic two-state fraction does. Moreover, the simulated subsurface enrichment is weaker than in the theory, whose local-pressure mapping overestimates it. Since the two free but unfitted parameters of the model change the interfacial fraction by tens of percent (SM, Sec.~S11), the magnitudes of the structural quantities are model dependent, while their signs and trends are not.

Chain length plays no role in this structure (Table~\ref{tab:struct}). In the theory, the alkane acts on water only through the water density profile, which is nearly the same for both alkanes (10--90\% widths 0.59--0.61\,nm). The simulations agree with this result. The fractions next to octane and heptane differ by no more than their uncertainty, whereas replacing the alkane by vapor changes the interfacial fraction by up to 0.08 with our region definitions (0.04 with those of the companion paper, its Table S15, which reaches the same conclusion from the populations and the hydrogen-bond dynamics of interfacial water\cite{companionMD}). Therefore, the structural balance of interfacial water cannot explain the different wetting of the two alkanes, which is carried by the alkane-side tensions.\cite{companionMD} Nor, in this theory, does it feed back on $\gamma_{aw}$, which is computed before the two-state analysis is applied.

Below 240\K, the structural trends change character [Fig.~\ref{fig:sum}]. The subsurface maximum of $\rho_s$ disappears, the $\rho$ state becomes enriched beneath the surface, the $s$-state front moves 0.4--0.5\,nm into the liquid, and $\Gamma_s$ drops to $-4.6$ to $-5.1\,\mathrm{nm}^{-2}$ at 230\K. This is because within a few kelvin of the LLCP of the EOS, the $s$-state fraction becomes extremely sensitive to pressure, and the local model, which extrapolates the EOS to the pressures of a stretched interface, is no longer reliable.\cite{Caupin2019} Indeed, the stability limit used to saturate the local pressure jumps from $-225$ to $-93$\,MPa between 239 and 238\K\ (SM, Sec.~S11). These results belong to the extrapolated regime of the EOS, and we do not build on them.

\section{Discussion}

We can now separate what is established, what is model dependent, and what is open.

\emph{Established within the theory and consistent with experiment.} With cross parameters fixed at one temperature, square-gradient PC-SAFT with a four-site water model reproduces the measured temperature dependence of the alkane--water tension between 293 and 343\K, within 5\% for heptane and octane and within 10\% for hexane and nonane. The two parameter sets that do so agree on the wetting temperatures within the uncertainty set by the anchor and slope of $\gamma_{aw}$, with values of 271--273 and 352--362\K. In the theory, the lens angle of heptane and octane levels off on cooling below about 250\K. Within the range where $\gamma_w$ is measured, and for the transmitted fraction of the theory, this leveling-off is fixed by the data to $\pm2^\circ$ (octane) and does not depend on the parameter set, the extrapolation form or the equation of state.

\emph{Model dependent.} Absolute wetting temperatures are the least robust output of both approaches. In theory and simulation alike, $T_w$ moves by 40--60\K\ per mN\,m$^{-1}$ of error in $S$. Therefore, the remaining differences between them, 15 and 51\K, amount to only 0.3 and 1\mNm, whereas the shape of $\theta_c(T)$ and the transmitted fraction depend on how the tensions change with temperature and are much less sensitive. The transmitted fraction $f\approx0.5$ follows from imposing the anomaly through $c_w$ alone, which gives $f=(\gamma_{aw}/\gamma_w)\bar w$, where $\bar w$ is the share of the oil--water gradient energy carried by the water gradient under the geometric-mean combining rule. Other ways of imposing the anomaly give values between 0.32 and 0.78 (Table~\ref{tab:fvar}). Therefore, the partial transmission itself does not depend on this choice, whereas its magnitude does. The simulations, which have no combining rule, do not resolve $f$. Real water may transmit the anomaly partly or fully, and whether it shows the leveling-off depends on $f$. For $f=1$, the angle keeps rising on cooling (Table~\ref{tab:probe}), which is why the angle near 240\K\ is a test. The depth and position of the contact-angle maximum below 240\K, the effect of the low-density expansion on $\gamma_{aw}$ (nil to 1.7\mNm\ across parameter sets, growing with the water--alkane dispersion energy) and the magnitudes of the interfacial two-state quantities are likewise model dependent, whereas their signs and trends are not.

\emph{Open, and decidable by experiment.} Two measurements would fix $f$ for real water. First, the alkane--water tension between 273 and 240\K, measured at a precision of 0.1\mNm, distinguishes $f=0$, 0.5 and 1 near 240\K\ (steps of about 0.25\mNm\ at 240\K\ and 0.7\mNm\ at 230\K, but only 0.1\mNm\ at 250\K). In a single $\gamma_{aw}(T)$ curve, the transmitted anomaly, which raises $\gamma_{aw}$ above the regular continuation of its stable-liquid behavior, and the expansion effect, which lowers it, are resolved only in their sum. Nevertheless, the sum discriminates between the descriptions, because at 230\K\ the main four-site set places $\gamma_{aw}$ about 0.75\mNm\ above the regular continuation, whereas the two-site set places it about 1.1\mNm\ below it. Second, in the theory, the contact angle of octane lenses on supercooled water between 250 and 240\K\ should stay near $32^\circ$. Octane remains liquid in this range, since it freezes at 216\K\ (heptane at 183\K). The value of the angle at 240\K, measured to $\pm1^\circ$, constrains $f$ to about $\pm0.35$ if $\gamma_w$ is known to the scatter-based precision of Sec.~\ref{sec:iapwsE}, and a turn-down below 240\K, if the supercooling can be reached, would signal $f<1$. In addition, the interfacial structure, with a disordered outer layer and tetrahedral water enriched 0.4--0.6\,nm below it, should appear in surface-specific vibrational spectroscopy as a tetrahedrally coordinated O--H signal that grows on heating and originates beneath the topmost layer.

\emph{Outlook.} A theory in which the surface-tension anomaly of water is derived rather than imposed would decide $f$ from first principles. The two-structure interfacial models of Hrub\'y and Holten\cite{HrubyHolten2004} and of Feeney and Debenedetti\cite{Feeney2003} did this for the free surface. Recently, a two-state formulation of PC-SAFT has become available.\cite{Novak2025} If it is used within square-gradient theory, with its own influence parameters for the two states, it would make the interfacial two-state balance thermodynamically self-consistent, and the transmission of the anomaly to the oil--water interface would become a prediction. In addition, classical density-functional versions of PC-SAFT\cite{Rehner2023} would remove the square-gradient approximation. Until then, the present results provide the experimental targets, namely a lens angle whose growth, if $f<1$, nearly stops below about 250\K, and an oil--water tension whose departure from regular behavior below 273\K\ measures how much of the surface anomaly of water is inherited by the oil--water interface.

\section{Conclusions}

In this work, we asked whether the surface-tension anomaly of supercooled water belongs to the liquid, and thus reaches every interface, or to the free surface alone. We examined this question for alkane--water interfaces using square-gradient PC-SAFT theory coupled to the two-state equation of state of real water. Three results stand out.

First, the theory is anchored in experiment wherever experimental data exist. With the measured surface tension of supercooled water as input and two cross parameters fixed at 295\K, it reproduces the measured slopes of the heptane-- and octane--water tensions between 293 and 343\K\ within 5\%. The same procedure reproduces the slopes for hexane and nonane within 10\%. In addition, the theory predicts wetting temperatures of $271\pm14$ and $362\pm20\K$ for heptane and octane, respectively.

Second, the anomaly is only partly transmitted. In the theory, 55\% of the anomalous rise of the water tension reaches the oil--water interface, nearly the same for both alkanes and at all temperatures, while the remaining 45\% enters the spreading coefficient. Therefore, the lens angle, which grows on cooling down to about 250\K, nearly stops growing below this temperature. In particular, octane lenses should stay near $32^\circ$ between 250 and 240\K. This fraction is a property of the interface, not of the anomaly. With the anomaly carried by the water influence parameter, it equals the tension ratio $\gamma_{aw}/\gamma_w$ times the share of the oil--water gradient energy carried by water. Other ways of imposing the anomaly give fractions between 0.32 and 0.78, which means that the transmission is partial in every variant that we examined. Below 241.7\K, the plateau also relies on the exponential continuation of the measured anomaly.

Third, theory and molecular dynamics simulations of TIP4P/2005 water agree on the structure of interfacial water. Both show a disordered outer layer over a tetrahedrally enriched subsurface, with a net adsorption of tetrahedral water that changes sign between about 285 and 320\K\ and does not depend on the alkane chain length. The simulations, however, do not resolve the transmitted fraction.

Therefore, whether real water transmits its anomaly partly or fully is an experimental question with a definite answer. For the fraction found in the theory, the octane--water tension at 240\K\ lies 0.3\mNm\ above the regular continuation of its stable-liquid behavior, which is a difference that a measurement to 0.1\mNm\ would resolve. Moreover, the octane lens angle at 240\K, measured to $\pm1^\circ$, would fix the fraction to about $\pm0.35$. Either measurement would show how much of the surface anomaly of water is inherited by an oil--water interface.

\section*{Supplementary Material}

See the supplementary material for the complete equations and parameters of the two-state EOS and of PC-SAFT, the numerical method of the square-gradient calculation and its convergence, the IAPWS-E fit, tabulated tensions, spreading coefficients and contact angles, the parameter-set comparison, the two control calculations, the robustness analysis, the local two-state model and its sensitivity, the calculations for hexane and nonane, the transmitted fraction under alternative assumptions, the tabulated interfacial structure from theory and simulation, and the derivatives of the simulated tensions.

\begin{acknowledgments}
The analysis of the simulation trajectories was performed on the Mill HPC cluster at Missouri University of Science and Technology, supported by the National Science Foundation under Grant No.~OAC-1919789.
\end{acknowledgments}

\section*{Author declarations}
\subsection*{Conflict of Interest}
The authors have no conflicts to disclose.
\subsection*{Author Contributions}
\textbf{Fawaz Hrahsheh:} Conceptualization (lead); Formal analysis (lead); Investigation (lead); Methodology (lead); Software (lead); Visualization (lead); Writing -- original draft (lead); Writing -- review \& editing (equal). \textbf{Moaz M. Altarawneh:} Formal analysis (supporting); Validation (supporting); Writing -- review \& editing (equal). \textbf{Gerald Wilemski:} Conceptualization (supporting); Supervision (lead); Writing -- review \& editing (equal).

\section*{Data availability}
The data that support the findings of this study are available within the article and its supplementary material. The Python code of the two-state, PC-SAFT and square-gradient calculations and the computed tensions and profiles are available from the corresponding author upon reasonable request.

\bibliography{refs}

\begin{thebibliography}{54}%
\makeatletter
\providecommand \@ifxundefined [1]{%
 \@ifx{#1\undefined}
}%
\providecommand \@ifnum [1]{%
 \ifnum #1\expandafter \@firstoftwo
 \else \expandafter \@secondoftwo
 \fi
}%
\providecommand \@ifx [1]{%
 \ifx #1\expandafter \@firstoftwo
 \else \expandafter \@secondoftwo
 \fi
}%
\providecommand \natexlab [1]{#1}%
\providecommand \enquote  [1]{``#1''}%
\providecommand \bibnamefont  [1]{#1}%
\providecommand \bibfnamefont [1]{#1}%
\providecommand \citenamefont [1]{#1}%
\providecommand \href@noop [0]{\@secondoftwo}%
\providecommand \href [0]{\begingroup \@sanitize@url \@href}%
\providecommand \@href[1]{\@@startlink{#1}\@@href}%
\providecommand \@@href[1]{\endgroup#1\@@endlink}%
\providecommand \@sanitize@url [0]{\catcode `\\12\catcode `\$12\catcode
  `\&12\catcode `\#12\catcode `\^12\catcode `\_12\catcode `\%12\relax}%
\providecommand \@@startlink[1]{}%
\providecommand \@@endlink[0]{}%
\providecommand \url  [0]{\begingroup\@sanitize@url \@url }%
\providecommand \@url [1]{\endgroup\@href {#1}{\urlprefix }}%
\providecommand \urlprefix  [0]{URL }%
\providecommand \Eprint [0]{\href }%
\providecommand \doibase [0]{https://doi.org/}%
\providecommand \selectlanguage [0]{\@gobble}%
\providecommand \bibinfo  [0]{\@secondoftwo}%
\providecommand \bibfield  [0]{\@secondoftwo}%
\providecommand \translation [1]{[#1]}%
\providecommand \BibitemOpen [0]{}%
\providecommand \bibitemStop [0]{}%
\providecommand \bibitemNoStop [0]{.\EOS\space}%
\providecommand \EOS [0]{\spacefactor3000\relax}%
\providecommand \BibitemShut  [1]{\csname bibitem#1\endcsname}%
\let\auto@bib@innerbib\@empty
\bibitem [{\citenamefont {Tanaka}(2000)}]{Tanaka2000}%
  \BibitemOpen
  \bibfield  {author} {\bibinfo {author} {\bibfnamefont {H.}~\bibnamefont
  {Tanaka}},\ }\bibfield  {title} {\enquote {\bibinfo {title} {Simple physical
  model of liquid water},}\ }\href@noop {} {\bibfield  {journal} {\bibinfo
  {journal} {J. Chem. Phys.}\ }\textbf {\bibinfo {volume} {112}},\ \bibinfo
  {pages} {799} (\bibinfo {year} {2000})}\BibitemShut {NoStop}%
\bibitem [{\citenamefont {Russo}\ and\ \citenamefont
  {Tanaka}(2014)}]{Russo2014}%
  \BibitemOpen
  \bibfield  {author} {\bibinfo {author} {\bibfnamefont {J.}~\bibnamefont
  {Russo}}\ and\ \bibinfo {author} {\bibfnamefont {H.}~\bibnamefont {Tanaka}},\
  }\bibfield  {title} {\enquote {\bibinfo {title} {Understanding water's
  anomalies with locally favoured structures},}\ }\href@noop {} {\bibfield
  {journal} {\bibinfo  {journal} {Nat. Commun.}\ }\textbf {\bibinfo {volume}
  {5}},\ \bibinfo {pages} {3556} (\bibinfo {year} {2014})}\BibitemShut
  {NoStop}%
\bibitem [{\citenamefont {Gallo}\ \emph {et~al.}(2016)\citenamefont {Gallo},
  \citenamefont {Amann-Winkel}, \citenamefont {Angell}, \citenamefont
  {Anisimov}, \citenamefont {Caupin}, \citenamefont {Chakravarty},
  \citenamefont {Lascaris}, \citenamefont {Loerting}, \citenamefont
  {Panagiotopoulos}, \citenamefont {Russo}, \citenamefont {Sellberg},
  \citenamefont {Stanley}, \citenamefont {Tanaka}, \citenamefont {Vega},
  \citenamefont {Xu},\ and\ \citenamefont {Pettersson}}]{Gallo2016}%
  \BibitemOpen
  \bibfield  {author} {\bibinfo {author} {\bibfnamefont {P.}~\bibnamefont
  {Gallo}}, \bibinfo {author} {\bibfnamefont {K.}~\bibnamefont {Amann-Winkel}},
  \bibinfo {author} {\bibfnamefont {C.~A.}\ \bibnamefont {Angell}}, \bibinfo
  {author} {\bibfnamefont {M.~A.}\ \bibnamefont {Anisimov}}, \bibinfo {author}
  {\bibfnamefont {F.}~\bibnamefont {Caupin}}, \bibinfo {author} {\bibfnamefont
  {C.}~\bibnamefont {Chakravarty}}, \bibinfo {author} {\bibfnamefont
  {E.}~\bibnamefont {Lascaris}}, \bibinfo {author} {\bibfnamefont
  {T.}~\bibnamefont {Loerting}}, \bibinfo {author} {\bibfnamefont {A.~Z.}\
  \bibnamefont {Panagiotopoulos}}, \bibinfo {author} {\bibfnamefont
  {J.}~\bibnamefont {Russo}}, \bibinfo {author} {\bibfnamefont {J.~A.}\
  \bibnamefont {Sellberg}}, \bibinfo {author} {\bibfnamefont {H.~E.}\
  \bibnamefont {Stanley}}, \bibinfo {author} {\bibfnamefont {H.}~\bibnamefont
  {Tanaka}}, \bibinfo {author} {\bibfnamefont {C.}~\bibnamefont {Vega}},
  \bibinfo {author} {\bibfnamefont {L.}~\bibnamefont {Xu}},\ and\ \bibinfo
  {author} {\bibfnamefont {L.~G.~M.}\ \bibnamefont {Pettersson}},\ }\bibfield
  {title} {\enquote {\bibinfo {title} {Water: A tale of two liquids},}\
  }\href@noop {} {\bibfield  {journal} {\bibinfo  {journal} {Chem. Rev.}\
  }\textbf {\bibinfo {volume} {116}},\ \bibinfo {pages} {7463} (\bibinfo {year}
  {2016})}\BibitemShut {NoStop}%
\bibitem [{\citenamefont {Anisimov}\ \emph {et~al.}(2018)\citenamefont
  {Anisimov}, \citenamefont {Duska}, \citenamefont {Caupin}, \citenamefont
  {Amrhein}, \citenamefont {Rosenbaum},\ and\ \citenamefont
  {Sadus}}]{Anisimov2018}%
  \BibitemOpen
  \bibfield  {author} {\bibinfo {author} {\bibfnamefont {M.~A.}\ \bibnamefont
  {Anisimov}}, \bibinfo {author} {\bibfnamefont {M.}~\bibnamefont {Duska}},
  \bibinfo {author} {\bibfnamefont {F.}~\bibnamefont {Caupin}}, \bibinfo
  {author} {\bibfnamefont {L.~E.}\ \bibnamefont {Amrhein}}, \bibinfo {author}
  {\bibfnamefont {A.}~\bibnamefont {Rosenbaum}},\ and\ \bibinfo {author}
  {\bibfnamefont {R.~J.}\ \bibnamefont {Sadus}},\ }\bibfield  {title} {\enquote
  {\bibinfo {title} {Thermodynamics of fluid polyamorphism},}\ }\href@noop {}
  {\bibfield  {journal} {\bibinfo  {journal} {Phys. Rev. X}\ }\textbf {\bibinfo
  {volume} {8}},\ \bibinfo {pages} {011004} (\bibinfo {year}
  {2018})}\BibitemShut {NoStop}%
\bibitem [{\citenamefont {Poole}\ \emph {et~al.}(1992)\citenamefont {Poole},
  \citenamefont {Sciortino}, \citenamefont {Essmann},\ and\ \citenamefont
  {Stanley}}]{Poole1992}%
  \BibitemOpen
  \bibfield  {author} {\bibinfo {author} {\bibfnamefont {P.~H.}\ \bibnamefont
  {Poole}}, \bibinfo {author} {\bibfnamefont {F.}~\bibnamefont {Sciortino}},
  \bibinfo {author} {\bibfnamefont {U.}~\bibnamefont {Essmann}},\ and\ \bibinfo
  {author} {\bibfnamefont {H.~E.}\ \bibnamefont {Stanley}},\ }\bibfield
  {title} {\enquote {\bibinfo {title} {Phase behaviour of metastable water},}\
  }\href@noop {} {\bibfield  {journal} {\bibinfo  {journal} {Nature}\ }\textbf
  {\bibinfo {volume} {360}},\ \bibinfo {pages} {324} (\bibinfo {year}
  {1992})}\BibitemShut {NoStop}%
\bibitem [{\citenamefont {Holten}, \citenamefont {Sengers},\ and\ \citenamefont
  {Anisimov}(2014)}]{Holten2014}%
  \BibitemOpen
  \bibfield  {author} {\bibinfo {author} {\bibfnamefont {V.}~\bibnamefont
  {Holten}}, \bibinfo {author} {\bibfnamefont {J.~V.}\ \bibnamefont
  {Sengers}},\ and\ \bibinfo {author} {\bibfnamefont {M.~A.}\ \bibnamefont
  {Anisimov}},\ }\bibfield  {title} {\enquote {\bibinfo {title} {Equation of
  state for supercooled water at pressures up to 400 {MPa}},}\ }\href@noop {}
  {\bibfield  {journal} {\bibinfo  {journal} {J. Phys. Chem. Ref. Data}\
  }\textbf {\bibinfo {volume} {43}},\ \bibinfo {pages} {043101} (\bibinfo
  {year} {2014})}\BibitemShut {NoStop}%
\bibitem [{\citenamefont {Debenedetti}, \citenamefont {Sciortino},\ and\
  \citenamefont {Zerze}(2020)}]{Debenedetti2020}%
  \BibitemOpen
  \bibfield  {author} {\bibinfo {author} {\bibfnamefont {P.~G.}\ \bibnamefont
  {Debenedetti}}, \bibinfo {author} {\bibfnamefont {F.}~\bibnamefont
  {Sciortino}},\ and\ \bibinfo {author} {\bibfnamefont {G.~H.}\ \bibnamefont
  {Zerze}},\ }\bibfield  {title} {\enquote {\bibinfo {title} {Second critical
  point in two realistic models of water},}\ }\href@noop {} {\bibfield
  {journal} {\bibinfo  {journal} {Science}\ }\textbf {\bibinfo {volume}
  {369}},\ \bibinfo {pages} {289} (\bibinfo {year} {2020})}\BibitemShut
  {NoStop}%
\bibitem [{\citenamefont {Kim}\ \emph {et~al.}(2020)\citenamefont {Kim},
  \citenamefont {Amann-Winkel}, \citenamefont {Giovambattista}, \citenamefont
  {Sp{\"a}h}, \citenamefont {Perakis}, \citenamefont {Pathak}, \citenamefont
  {{Ladd Parada}}, \citenamefont {Yang}, \citenamefont {Mariedahl},
  \citenamefont {Eklund}, \citenamefont {Lane}, \citenamefont {You},
  \citenamefont {Jeong}, \citenamefont {Weston}, \citenamefont {Lee},
  \citenamefont {Eom}, \citenamefont {Kim}, \citenamefont {Park}, \citenamefont
  {Chun}, \citenamefont {Poole},\ and\ \citenamefont {Nilsson}}]{Kim2020}%
  \BibitemOpen
  \bibfield  {author} {\bibinfo {author} {\bibfnamefont {K.~H.}\ \bibnamefont
  {Kim}}, \bibinfo {author} {\bibfnamefont {K.}~\bibnamefont {Amann-Winkel}},
  \bibinfo {author} {\bibfnamefont {N.}~\bibnamefont {Giovambattista}},
  \bibinfo {author} {\bibfnamefont {A.}~\bibnamefont {Sp{\"a}h}}, \bibinfo
  {author} {\bibfnamefont {F.}~\bibnamefont {Perakis}}, \bibinfo {author}
  {\bibfnamefont {H.}~\bibnamefont {Pathak}}, \bibinfo {author} {\bibfnamefont
  {M.}~\bibnamefont {{Ladd Parada}}}, \bibinfo {author} {\bibfnamefont
  {C.}~\bibnamefont {Yang}}, \bibinfo {author} {\bibfnamefont {D.}~\bibnamefont
  {Mariedahl}}, \bibinfo {author} {\bibfnamefont {T.}~\bibnamefont {Eklund}},
  \bibinfo {author} {\bibfnamefont {T.~J.}\ \bibnamefont {Lane}}, \bibinfo
  {author} {\bibfnamefont {S.}~\bibnamefont {You}}, \bibinfo {author}
  {\bibfnamefont {S.}~\bibnamefont {Jeong}}, \bibinfo {author} {\bibfnamefont
  {M.}~\bibnamefont {Weston}}, \bibinfo {author} {\bibfnamefont {J.~H.}\
  \bibnamefont {Lee}}, \bibinfo {author} {\bibfnamefont {I.}~\bibnamefont
  {Eom}}, \bibinfo {author} {\bibfnamefont {M.}~\bibnamefont {Kim}}, \bibinfo
  {author} {\bibfnamefont {J.}~\bibnamefont {Park}}, \bibinfo {author}
  {\bibfnamefont {S.~H.}\ \bibnamefont {Chun}}, \bibinfo {author}
  {\bibfnamefont {P.~H.}\ \bibnamefont {Poole}},\ and\ \bibinfo {author}
  {\bibfnamefont {A.}~\bibnamefont {Nilsson}},\ }\bibfield  {title} {\enquote
  {\bibinfo {title} {Experimental observation of the liquid-liquid transition
  in bulk supercooled water under pressure},}\ }\href@noop {} {\bibfield
  {journal} {\bibinfo  {journal} {Science}\ }\textbf {\bibinfo {volume}
  {370}},\ \bibinfo {pages} {978} (\bibinfo {year} {2020})}\BibitemShut
  {NoStop}%
\bibitem [{\citenamefont {{IAPWS}}(2015)}]{IAPWSG12}%
  \BibitemOpen
  \bibfield  {author} {\bibinfo {author} {\bibnamefont {{IAPWS}}},\ }\href@noop
  {} {\enquote {\bibinfo {title} {Guideline on thermodynamic properties of
  supercooled water, {IAPWS G12-15}},}\ }\bibinfo {howpublished} {International
  Association for the Properties of Water and Steam} (\bibinfo {year}
  {2015})\BibitemShut {NoStop}%
\bibitem [{\citenamefont {Floriano}\ and\ \citenamefont
  {Angell}(1990)}]{Floriano1990}%
  \BibitemOpen
  \bibfield  {author} {\bibinfo {author} {\bibfnamefont {M.~A.}\ \bibnamefont
  {Floriano}}\ and\ \bibinfo {author} {\bibfnamefont {C.~A.}\ \bibnamefont
  {Angell}},\ }\bibfield  {title} {\enquote {\bibinfo {title} {Surface tension
  and molar surface free energy and entropy of water to $-27.2\,^\circ${C}},}\
  }\href@noop {} {\bibfield  {journal} {\bibinfo  {journal} {J. Phys. Chem.}\
  }\textbf {\bibinfo {volume} {94}},\ \bibinfo {pages} {4199} (\bibinfo {year}
  {1990})}\BibitemShut {NoStop}%
\bibitem [{\citenamefont {Hrub{\'y}}\ \emph {et~al.}(2014)\citenamefont
  {Hrub{\'y}}, \citenamefont {Vin{\v{s}}}, \citenamefont {Mare{\v{s}}},
  \citenamefont {Hykl},\ and\ \citenamefont {Kalov{\'a}}}]{Hruby2014}%
  \BibitemOpen
  \bibfield  {author} {\bibinfo {author} {\bibfnamefont {J.}~\bibnamefont
  {Hrub{\'y}}}, \bibinfo {author} {\bibfnamefont {V.}~\bibnamefont
  {Vin{\v{s}}}}, \bibinfo {author} {\bibfnamefont {R.}~\bibnamefont
  {Mare{\v{s}}}}, \bibinfo {author} {\bibfnamefont {J.}~\bibnamefont {Hykl}},\
  and\ \bibinfo {author} {\bibfnamefont {J.}~\bibnamefont {Kalov{\'a}}},\
  }\bibfield  {title} {\enquote {\bibinfo {title} {Surface tension of
  supercooled water: No inflection point down to $-25$~$^\circ${C}},}\
  }\href@noop {} {\bibfield  {journal} {\bibinfo  {journal} {J. Phys. Chem.
  Lett.}\ }\textbf {\bibinfo {volume} {5}},\ \bibinfo {pages} {425} (\bibinfo
  {year} {2014})}\BibitemShut {NoStop}%
\bibitem [{\citenamefont {Vin{\v{s}}}\ \emph {et~al.}(2015)\citenamefont
  {Vin{\v{s}}}, \citenamefont {Fransen}, \citenamefont {Hykl},\ and\
  \citenamefont {Hrub{\'y}}}]{Vins2015}%
  \BibitemOpen
  \bibfield  {author} {\bibinfo {author} {\bibfnamefont {V.}~\bibnamefont
  {Vin{\v{s}}}}, \bibinfo {author} {\bibfnamefont {M.}~\bibnamefont {Fransen}},
  \bibinfo {author} {\bibfnamefont {J.}~\bibnamefont {Hykl}},\ and\ \bibinfo
  {author} {\bibfnamefont {J.}~\bibnamefont {Hrub{\'y}}},\ }\bibfield  {title}
  {\enquote {\bibinfo {title} {Surface tension of supercooled water determined
  by using a counterpressure capillary rise method},}\ }\href@noop {}
  {\bibfield  {journal} {\bibinfo  {journal} {J. Phys. Chem. B}\ }\textbf
  {\bibinfo {volume} {119}},\ \bibinfo {pages} {5567} (\bibinfo {year}
  {2015})}\BibitemShut {NoStop}%
\bibitem [{\citenamefont {Vin{\v{s}}}\ \emph {et~al.}(2020)\citenamefont
  {Vin{\v{s}}}, \citenamefont {Hykl}, \citenamefont {Hrub{\'y}}, \citenamefont
  {Blahut}, \citenamefont {Celn{\'y}}, \citenamefont {{\v{C}}ensk{\'y}},\ and\
  \citenamefont {Prokopov{\'a}}}]{Vins2020}%
  \BibitemOpen
  \bibfield  {author} {\bibinfo {author} {\bibfnamefont {V.}~\bibnamefont
  {Vin{\v{s}}}}, \bibinfo {author} {\bibfnamefont {J.}~\bibnamefont {Hykl}},
  \bibinfo {author} {\bibfnamefont {J.}~\bibnamefont {Hrub{\'y}}}, \bibinfo
  {author} {\bibfnamefont {A.}~\bibnamefont {Blahut}}, \bibinfo {author}
  {\bibfnamefont {D.}~\bibnamefont {Celn{\'y}}}, \bibinfo {author}
  {\bibfnamefont {M.}~\bibnamefont {{\v{C}}ensk{\'y}}},\ and\ \bibinfo {author}
  {\bibfnamefont {O.}~\bibnamefont {Prokopov{\'a}}},\ }\bibfield  {title}
  {\enquote {\bibinfo {title} {Possible anomaly in the surface tension of
  supercooled water: New experiments at extreme supercooling down to
  $-31.4$~$^\circ${C}},}\ }\href@noop {} {\bibfield  {journal} {\bibinfo
  {journal} {J. Phys. Chem. Lett.}\ }\textbf {\bibinfo {volume} {11}},\
  \bibinfo {pages} {4443} (\bibinfo {year} {2020})}\BibitemShut {NoStop}%
\bibitem [{\citenamefont {Hrub\'y}\ and\ \citenamefont
  {Holten}(2004)}]{HrubyHolten2004}%
  \BibitemOpen
  \bibfield  {author} {\bibinfo {author} {\bibfnamefont {J.}~\bibnamefont
  {Hrub\'y}}\ and\ \bibinfo {author} {\bibfnamefont {V.}~\bibnamefont
  {Holten}},\ }\bibfield  {title} {\enquote {\bibinfo {title} {A two-structure
  model of thermodynamic properties and surface tension of supercooled
  water},}\ }\href@noop {} {\bibfield  {journal} {\bibinfo  {journal} {Int. J.
  Thermophys.}\ }\textbf {\bibinfo {volume} {25}},\ \bibinfo {pages} {1135}
  (\bibinfo {year} {2004})}\BibitemShut {NoStop}%
\bibitem [{\citenamefont {Feeney}\ and\ \citenamefont
  {Debenedetti}(2003)}]{Feeney2003}%
  \BibitemOpen
  \bibfield  {author} {\bibinfo {author} {\bibfnamefont {M.~E.}\ \bibnamefont
  {Feeney}}\ and\ \bibinfo {author} {\bibfnamefont {P.~G.}\ \bibnamefont
  {Debenedetti}},\ }\bibfield  {title} {\enquote {\bibinfo {title} {A
  theoretical study of the interfacial properties of supercooled water},}\
  }\href@noop {} {\bibfield  {journal} {\bibinfo  {journal} {Ind. Eng. Chem.
  Res.}\ }\textbf {\bibinfo {volume} {42}},\ \bibinfo {pages} {6396} (\bibinfo
  {year} {2003})}\BibitemShut {NoStop}%
\bibitem [{\citenamefont {Longo}\ \emph {et~al.}(2023)\citenamefont {Longo},
  \citenamefont {Buldyrev}, \citenamefont {Anisimov},\ and\ \citenamefont
  {Caupin}}]{Longo2023}%
  \BibitemOpen
  \bibfield  {author} {\bibinfo {author} {\bibfnamefont {T.~J.}\ \bibnamefont
  {Longo}}, \bibinfo {author} {\bibfnamefont {S.~V.}\ \bibnamefont {Buldyrev}},
  \bibinfo {author} {\bibfnamefont {M.~A.}\ \bibnamefont {Anisimov}},\ and\
  \bibinfo {author} {\bibfnamefont {F.}~\bibnamefont {Caupin}},\ }\bibfield
  {title} {\enquote {\bibinfo {title} {Interfacial properties of fluids
  exhibiting liquid polyamorphism and water-like anomalies},}\ }\href@noop {}
  {\bibfield  {journal} {\bibinfo  {journal} {J. Phys. Chem. B}\ }\textbf
  {\bibinfo {volume} {127}},\ \bibinfo {pages} {3079} (\bibinfo {year}
  {2023})}\BibitemShut {NoStop}%
\bibitem [{\citenamefont {Malek}, \citenamefont {Poole},\ and\ \citenamefont
  {Saika-Voivod}(2019)}]{Malek2019}%
  \BibitemOpen
  \bibfield  {author} {\bibinfo {author} {\bibfnamefont {S.~M.~A.}\
  \bibnamefont {Malek}}, \bibinfo {author} {\bibfnamefont {P.~H.}\ \bibnamefont
  {Poole}},\ and\ \bibinfo {author} {\bibfnamefont {I.}~\bibnamefont
  {Saika-Voivod}},\ }\bibfield  {title} {\enquote {\bibinfo {title} {Surface
  tension of supercooled water nanodroplets from computer simulations},}\
  }\href@noop {} {\bibfield  {journal} {\bibinfo  {journal} {J. Chem. Phys.}\
  }\textbf {\bibinfo {volume} {150}},\ \bibinfo {pages} {234507} (\bibinfo
  {year} {2019})}\BibitemShut {NoStop}%
\bibitem [{\citenamefont {Hrahsheh}, \citenamefont {Jum'h},\ and\ \citenamefont
  {Wilemski}(2024)}]{Hrahsheh2024}%
  \BibitemOpen
  \bibfield  {author} {\bibinfo {author} {\bibfnamefont {F.}~\bibnamefont
  {Hrahsheh}}, \bibinfo {author} {\bibfnamefont {I.}~\bibnamefont {Jum'h}},\
  and\ \bibinfo {author} {\bibfnamefont {G.}~\bibnamefont {Wilemski}},\
  }\bibfield  {title} {\enquote {\bibinfo {title} {Second inflection point of
  supercooled water surface tension induced by hydrogen bonds: A
  molecular-dynamics study},}\ }\href@noop {} {\bibfield  {journal} {\bibinfo
  {journal} {J. Chem. Phys.}\ }\textbf {\bibinfo {volume} {160}},\ \bibinfo
  {pages} {114504} (\bibinfo {year} {2024})}\BibitemShut {NoStop}%
\bibitem [{\citenamefont {Gorfer}, \citenamefont {Dellago},\ and\ \citenamefont
  {Sega}(2023)}]{Gorfer2023}%
  \BibitemOpen
  \bibfield  {author} {\bibinfo {author} {\bibfnamefont {A.}~\bibnamefont
  {Gorfer}}, \bibinfo {author} {\bibfnamefont {C.}~\bibnamefont {Dellago}},\
  and\ \bibinfo {author} {\bibfnamefont {M.}~\bibnamefont {Sega}},\ }\bibfield
  {title} {\enquote {\bibinfo {title} {High-density liquid ({HDL}) adsorption
  at the supercooled water/vapor interface and its possible relation to the
  second surface tension inflection point},}\ }\href@noop {} {\bibfield
  {journal} {\bibinfo  {journal} {J. Chem. Phys.}\ }\textbf {\bibinfo {volume}
  {158}},\ \bibinfo {pages} {054503} (\bibinfo {year} {2023})}\BibitemShut
  {NoStop}%
\bibitem [{\citenamefont {Yuan}\ \emph {et~al.}(2026)\citenamefont {Yuan},
  \citenamefont {Qiu}, \citenamefont {Sun},\ and\ \citenamefont
  {Tanaka}}]{Yuan2026}%
  \BibitemOpen
  \bibfield  {author} {\bibinfo {author} {\bibfnamefont {J.}~\bibnamefont
  {Yuan}}, \bibinfo {author} {\bibfnamefont {K.}~\bibnamefont {Qiu}}, \bibinfo
  {author} {\bibfnamefont {G.}~\bibnamefont {Sun}},\ and\ \bibinfo {author}
  {\bibfnamefont {H.}~\bibnamefont {Tanaka}},\ }\bibfield  {title} {\enquote
  {\bibinfo {title} {Competing hydrogen-bond orders drive water's anomalous
  surface tension},}\ }\href@noop {} {\bibfield  {journal} {\bibinfo  {journal}
  {Nat. Commun.}\ }\textbf {\bibinfo {volume} {17}},\ \bibinfo {pages} {1498}
  (\bibinfo {year} {2026})}\BibitemShut {NoStop}%
\bibitem [{\citenamefont {Ragil}\ \emph
  {et~al.}(1996{\natexlab{a}})\citenamefont {Ragil}, \citenamefont {Meunier},
  \citenamefont {Broseta}, \citenamefont {Indekeu},\ and\ \citenamefont
  {Bonn}}]{Ragil1996}%
  \BibitemOpen
  \bibfield  {author} {\bibinfo {author} {\bibfnamefont {K.}~\bibnamefont
  {Ragil}}, \bibinfo {author} {\bibfnamefont {J.}~\bibnamefont {Meunier}},
  \bibinfo {author} {\bibfnamefont {D.}~\bibnamefont {Broseta}}, \bibinfo
  {author} {\bibfnamefont {J.~O.}\ \bibnamefont {Indekeu}},\ and\ \bibinfo
  {author} {\bibfnamefont {D.}~\bibnamefont {Bonn}},\ }\bibfield  {title}
  {\enquote {\bibinfo {title} {Experimental observation of critical wetting},}\
  }\href@noop {} {\bibfield  {journal} {\bibinfo  {journal} {Phys. Rev. Lett.}\
  }\textbf {\bibinfo {volume} {77}},\ \bibinfo {pages} {1532} (\bibinfo {year}
  {1996}{\natexlab{a}})}\BibitemShut {NoStop}%
\bibitem [{\citenamefont {Shahidzadeh}\ \emph {et~al.}(1998)\citenamefont
  {Shahidzadeh}, \citenamefont {Bonn}, \citenamefont {Ragil}, \citenamefont
  {Broseta},\ and\ \citenamefont {Meunier}}]{Shahidzadeh1998}%
  \BibitemOpen
  \bibfield  {author} {\bibinfo {author} {\bibfnamefont {N.}~\bibnamefont
  {Shahidzadeh}}, \bibinfo {author} {\bibfnamefont {D.}~\bibnamefont {Bonn}},
  \bibinfo {author} {\bibfnamefont {K.}~\bibnamefont {Ragil}}, \bibinfo
  {author} {\bibfnamefont {D.}~\bibnamefont {Broseta}},\ and\ \bibinfo {author}
  {\bibfnamefont {J.}~\bibnamefont {Meunier}},\ }\bibfield  {title} {\enquote
  {\bibinfo {title} {Sequence of two wetting transitions induced by tuning the
  {H}amaker constant},}\ }\href@noop {} {\bibfield  {journal} {\bibinfo
  {journal} {Phys. Rev. Lett.}\ }\textbf {\bibinfo {volume} {80}},\ \bibinfo
  {pages} {3992} (\bibinfo {year} {1998})}\BibitemShut {NoStop}%
\bibitem [{\citenamefont {Pfohl}\ and\ \citenamefont
  {Riegler}(1999)}]{Pfohl1999}%
  \BibitemOpen
  \bibfield  {author} {\bibinfo {author} {\bibfnamefont {T.}~\bibnamefont
  {Pfohl}}\ and\ \bibinfo {author} {\bibfnamefont {H.}~\bibnamefont
  {Riegler}},\ }\bibfield  {title} {\enquote {\bibinfo {title} {Critical
  wetting of a liquid/vapor interface by octane},}\ }\href@noop {} {\bibfield
  {journal} {\bibinfo  {journal} {Phys. Rev. Lett.}\ }\textbf {\bibinfo
  {volume} {82}},\ \bibinfo {pages} {783} (\bibinfo {year} {1999})}\BibitemShut
  {NoStop}%
\bibitem [{\citenamefont {Bertrand}\ \emph {et~al.}(2000)\citenamefont
  {Bertrand}, \citenamefont {Dobbs}, \citenamefont {Broseta}, \citenamefont
  {Indekeu}, \citenamefont {Bonn},\ and\ \citenamefont
  {Meunier}}]{Bertrand2000}%
  \BibitemOpen
  \bibfield  {author} {\bibinfo {author} {\bibfnamefont {E.}~\bibnamefont
  {Bertrand}}, \bibinfo {author} {\bibfnamefont {H.}~\bibnamefont {Dobbs}},
  \bibinfo {author} {\bibfnamefont {D.}~\bibnamefont {Broseta}}, \bibinfo
  {author} {\bibfnamefont {J.}~\bibnamefont {Indekeu}}, \bibinfo {author}
  {\bibfnamefont {D.}~\bibnamefont {Bonn}},\ and\ \bibinfo {author}
  {\bibfnamefont {J.}~\bibnamefont {Meunier}},\ }\bibfield  {title} {\enquote
  {\bibinfo {title} {First-order and critical wetting of alkanes on water},}\
  }\href@noop {} {\bibfield  {journal} {\bibinfo  {journal} {Phys. Rev. Lett.}\
  }\textbf {\bibinfo {volume} {85}},\ \bibinfo {pages} {1282} (\bibinfo {year}
  {2000})}\BibitemShut {NoStop}%
\bibitem [{\citenamefont {Hrahsheh}\ and\ \citenamefont
  {Wilemski}(2023)}]{Hrahsheh2023}%
  \BibitemOpen
  \bibfield  {author} {\bibinfo {author} {\bibfnamefont {F.}~\bibnamefont
  {Hrahsheh}}\ and\ \bibinfo {author} {\bibfnamefont {G.}~\bibnamefont
  {Wilemski}},\ }\bibfield  {title} {\enquote {\bibinfo {title} {Effects of
  molecular size and orientation on the interfacial properties and wetting
  behavior of water/n-alkane systems: a molecular-dynamics study},}\
  }\href@noop {} {\bibfield  {journal} {\bibinfo  {journal} {Phys. Chem. Chem.
  Phys.}\ }\textbf {\bibinfo {volume} {25}},\ \bibinfo {pages} {5808} (\bibinfo
  {year} {2023})}\BibitemShut {NoStop}%
\bibitem [{\citenamefont {Hrahsheh}, \citenamefont {Altarawneh},\ and\
  \citenamefont {Wilemski}(2026)}]{companionMD}%
  \BibitemOpen
  \bibfield  {author} {\bibinfo {author} {\bibfnamefont {F.}~\bibnamefont
  {Hrahsheh}}, \bibinfo {author} {\bibfnamefont {M.~M.}\ \bibnamefont
  {Altarawneh}},\ and\ \bibinfo {author} {\bibfnamefont {G.}~\bibnamefont
  {Wilemski}},\ }\href@noop {} {\enquote {\bibinfo {title} {Sensitivity of
  \textit{n}-alkane wetting temperatures on water to interfacial tensions},}\ }
  (\bibinfo {year} {2026}),\ \bibinfo {note} {submitted to J. Chem.
  Phys.}\BibitemShut {Stop}%
\bibitem [{\citenamefont {Neupane}\ and\ \citenamefont
  {Wilemski}(2021)}]{Neupane2021}%
  \BibitemOpen
  \bibfield  {author} {\bibinfo {author} {\bibfnamefont {P.}~\bibnamefont
  {Neupane}}\ and\ \bibinfo {author} {\bibfnamefont {G.}~\bibnamefont
  {Wilemski}},\ }\bibfield  {title} {\enquote {\bibinfo {title} {Molecular
  dynamics study of wetting of alkanes on water: from high temperature to the
  supercooled region and the influence of second inflection points of
  interfacial tensions},}\ }\href@noop {} {\bibfield  {journal} {\bibinfo
  {journal} {Phys. Chem. Chem. Phys.}\ }\textbf {\bibinfo {volume} {23}},\
  \bibinfo {pages} {14465} (\bibinfo {year} {2021})}\BibitemShut {NoStop}%
\bibitem [{\citenamefont {Cahn}\ and\ \citenamefont
  {Hilliard}(1958)}]{Cahn1958}%
  \BibitemOpen
  \bibfield  {author} {\bibinfo {author} {\bibfnamefont {J.~W.}\ \bibnamefont
  {Cahn}}\ and\ \bibinfo {author} {\bibfnamefont {J.~E.}\ \bibnamefont
  {Hilliard}},\ }\bibfield  {title} {\enquote {\bibinfo {title} {Free energy of
  a nonuniform system. {I}. {I}nterfacial free energy},}\ }\href@noop {}
  {\bibfield  {journal} {\bibinfo  {journal} {J. Chem. Phys.}\ }\textbf
  {\bibinfo {volume} {28}},\ \bibinfo {pages} {258} (\bibinfo {year}
  {1958})}\BibitemShut {NoStop}%
\bibitem [{\citenamefont {Ragil}\ \emph
  {et~al.}(1996{\natexlab{b}})\citenamefont {Ragil}, \citenamefont {Bonn},
  \citenamefont {Broseta},\ and\ \citenamefont {Meunier}}]{RagilJCP1996}%
  \BibitemOpen
  \bibfield  {author} {\bibinfo {author} {\bibfnamefont {K.}~\bibnamefont
  {Ragil}}, \bibinfo {author} {\bibfnamefont {D.}~\bibnamefont {Bonn}},
  \bibinfo {author} {\bibfnamefont {D.}~\bibnamefont {Broseta}},\ and\ \bibinfo
  {author} {\bibfnamefont {J.}~\bibnamefont {Meunier}},\ }\bibfield  {title}
  {\enquote {\bibinfo {title} {Wetting of alkanes on water from a {C}ahn-type
  theory},}\ }\href@noop {} {\bibfield  {journal} {\bibinfo  {journal} {J.
  Chem. Phys.}\ }\textbf {\bibinfo {volume} {105}},\ \bibinfo {pages} {5160}
  (\bibinfo {year} {1996}{\natexlab{b}})}\BibitemShut {NoStop}%
\bibitem [{\citenamefont {Gross}\ and\ \citenamefont
  {Sadowski}(2001)}]{Gross2001}%
  \BibitemOpen
  \bibfield  {author} {\bibinfo {author} {\bibfnamefont {J.}~\bibnamefont
  {Gross}}\ and\ \bibinfo {author} {\bibfnamefont {G.}~\bibnamefont
  {Sadowski}},\ }\bibfield  {title} {\enquote {\bibinfo {title}
  {Perturbed-chain {SAFT}: An equation of state based on a perturbation theory
  for chain molecules},}\ }\href@noop {} {\bibfield  {journal} {\bibinfo
  {journal} {Ind. Eng. Chem. Res.}\ }\textbf {\bibinfo {volume} {40}},\
  \bibinfo {pages} {1244} (\bibinfo {year} {2001})}\BibitemShut {NoStop}%
\bibitem [{\citenamefont {Gross}\ and\ \citenamefont
  {Sadowski}(2002)}]{Gross2002}%
  \BibitemOpen
  \bibfield  {author} {\bibinfo {author} {\bibfnamefont {J.}~\bibnamefont
  {Gross}}\ and\ \bibinfo {author} {\bibfnamefont {G.}~\bibnamefont
  {Sadowski}},\ }\bibfield  {title} {\enquote {\bibinfo {title} {Application of
  the perturbed-chain {SAFT} equation of state to associating systems},}\
  }\href@noop {} {\bibfield  {journal} {\bibinfo  {journal} {Ind. Eng. Chem.
  Res.}\ }\textbf {\bibinfo {volume} {41}},\ \bibinfo {pages} {5510} (\bibinfo
  {year} {2002})}\BibitemShut {NoStop}%
\bibitem [{\citenamefont {Klink}, \citenamefont {Plankov{\'a}},\ and\
  \citenamefont {Gross}(2015)}]{Klink2015}%
  \BibitemOpen
  \bibfield  {author} {\bibinfo {author} {\bibfnamefont {C.}~\bibnamefont
  {Klink}}, \bibinfo {author} {\bibfnamefont {B.}~\bibnamefont
  {Plankov{\'a}}},\ and\ \bibinfo {author} {\bibfnamefont {J.}~\bibnamefont
  {Gross}},\ }\bibfield  {title} {\enquote {\bibinfo {title} {Density
  functional theory for liquid--liquid interfaces of mixtures using the
  perturbed-chain polar statistical associating fluid theory equation of
  state},}\ }\href@noop {} {\bibfield  {journal} {\bibinfo  {journal} {Ind.
  Eng. Chem. Res.}\ }\textbf {\bibinfo {volume} {54}},\ \bibinfo {pages} {4633}
  (\bibinfo {year} {2015})}\BibitemShut {NoStop}%
\bibitem [{\citenamefont {Hare}\ and\ \citenamefont
  {Sorensen}(1987)}]{Hare1987}%
  \BibitemOpen
  \bibfield  {author} {\bibinfo {author} {\bibfnamefont {D.~E.}\ \bibnamefont
  {Hare}}\ and\ \bibinfo {author} {\bibfnamefont {C.~M.}\ \bibnamefont
  {Sorensen}},\ }\bibfield  {title} {\enquote {\bibinfo {title} {The density of
  supercooled water. {II}. {B}ulk samples cooled to the homogeneous nucleation
  limit},}\ }\href@noop {} {\bibfield  {journal} {\bibinfo  {journal} {J. Chem.
  Phys.}\ }\textbf {\bibinfo {volume} {87}},\ \bibinfo {pages} {4840} (\bibinfo
  {year} {1987})}\BibitemShut {NoStop}%
\bibitem [{\citenamefont {Speedy}\ and\ \citenamefont
  {Angell}(1976)}]{Speedy1976}%
  \BibitemOpen
  \bibfield  {author} {\bibinfo {author} {\bibfnamefont {R.~J.}\ \bibnamefont
  {Speedy}}\ and\ \bibinfo {author} {\bibfnamefont {C.~A.}\ \bibnamefont
  {Angell}},\ }\bibfield  {title} {\enquote {\bibinfo {title} {Isothermal
  compressibility of supercooled water and evidence for a thermodynamic
  singularity at $-45$~$^\circ${C}},}\ }\href@noop {} {\bibfield  {journal}
  {\bibinfo  {journal} {J. Chem. Phys.}\ }\textbf {\bibinfo {volume} {65}},\
  \bibinfo {pages} {851} (\bibinfo {year} {1976})}\BibitemShut {NoStop}%
\bibitem [{\citenamefont {{IAPWS}}(2014)}]{IAPWSR1}%
  \BibitemOpen
  \bibfield  {author} {\bibinfo {author} {\bibnamefont {{IAPWS}}},\ }\href@noop
  {} {\enquote {\bibinfo {title} {Revised release on surface tension of
  ordinary water substance, {IAPWS R1-76(2014)}},}\ }\bibinfo {howpublished}
  {International Association for the Properties of Water and Steam} (\bibinfo
  {year} {2014})\BibitemShut {NoStop}%
\bibitem [{\citenamefont {Chapman}\ \emph {et~al.}(1990)\citenamefont
  {Chapman}, \citenamefont {Gubbins}, \citenamefont {Jackson},\ and\
  \citenamefont {Radosz}}]{Chapman1990}%
  \BibitemOpen
  \bibfield  {author} {\bibinfo {author} {\bibfnamefont {W.~G.}\ \bibnamefont
  {Chapman}}, \bibinfo {author} {\bibfnamefont {K.~E.}\ \bibnamefont
  {Gubbins}}, \bibinfo {author} {\bibfnamefont {G.}~\bibnamefont {Jackson}},\
  and\ \bibinfo {author} {\bibfnamefont {M.}~\bibnamefont {Radosz}},\
  }\bibfield  {title} {\enquote {\bibinfo {title} {New reference equation of
  state for associating liquids},}\ }\href@noop {} {\bibfield  {journal}
  {\bibinfo  {journal} {Ind. Eng. Chem. Res.}\ }\textbf {\bibinfo {volume}
  {29}},\ \bibinfo {pages} {1709} (\bibinfo {year} {1990})}\BibitemShut
  {NoStop}%
\bibitem [{\citenamefont {Diamantonis}\ and\ \citenamefont
  {Economou}(2011)}]{Diamantonis2011}%
  \BibitemOpen
  \bibfield  {author} {\bibinfo {author} {\bibfnamefont {N.~I.}\ \bibnamefont
  {Diamantonis}}\ and\ \bibinfo {author} {\bibfnamefont {I.~G.}\ \bibnamefont
  {Economou}},\ }\bibfield  {title} {\enquote {\bibinfo {title} {Evaluation of
  statistical associating fluid theory ({SAFT}) and perturbed chain-{SAFT}
  equations of state for the calculation of thermodynamic derivative properties
  of fluids related to carbon capture and sequestration},}\ }\href@noop {}
  {\bibfield  {journal} {\bibinfo  {journal} {Energy Fuels}\ }\textbf {\bibinfo
  {volume} {25}},\ \bibinfo {pages} {3334} (\bibinfo {year}
  {2011})}\BibitemShut {NoStop}%
\bibitem [{\citenamefont {Rehner}\ and\ \citenamefont
  {Gross}(2020)}]{Rehner2020}%
  \BibitemOpen
  \bibfield  {author} {\bibinfo {author} {\bibfnamefont {P.}~\bibnamefont
  {Rehner}}\ and\ \bibinfo {author} {\bibfnamefont {J.}~\bibnamefont {Gross}},\
  }\bibfield  {title} {\enquote {\bibinfo {title} {Multiobjective optimization
  of {PCP-SAFT} parameters for water and alcohols using surface tension
  data},}\ }\href@noop {} {\bibfield  {journal} {\bibinfo  {journal} {J. Chem.
  Eng. Data}\ }\textbf {\bibinfo {volume} {65}},\ \bibinfo {pages} {5698}
  (\bibinfo {year} {2020})}\BibitemShut {NoStop}%
\bibitem [{\citenamefont {Rom\'an-Ram\'irez}\ and\ \citenamefont
  {Leeke}(2016)}]{Roman2016}%
  \BibitemOpen
  \bibfield  {author} {\bibinfo {author} {\bibfnamefont {L.~A.}\ \bibnamefont
  {Rom\'an-Ram\'irez}}\ and\ \bibinfo {author} {\bibfnamefont {G.~A.}\
  \bibnamefont {Leeke}},\ }\bibfield  {title} {\enquote {\bibinfo {title}
  {Evaluation of the {PC-SAFT} equation of state for the prediction of the
  phase behavior of {VOCs} in supercritical carbon dioxide},}\ }\href@noop {}
  {\bibfield  {journal} {\bibinfo  {journal} {J. Chem. Eng. Data}\ }\textbf
  {\bibinfo {volume} {61}},\ \bibinfo {pages} {2777} (\bibinfo {year}
  {2016})}\BibitemShut {NoStop}%
\bibitem [{\citenamefont {Cameretti}\ and\ \citenamefont
  {Sadowski}(2008)}]{Cameretti2008}%
  \BibitemOpen
  \bibfield  {author} {\bibinfo {author} {\bibfnamefont {L.~F.}\ \bibnamefont
  {Cameretti}}\ and\ \bibinfo {author} {\bibfnamefont {G.}~\bibnamefont
  {Sadowski}},\ }\bibfield  {title} {\enquote {\bibinfo {title} {Modeling of
  aqueous amino acid and polypeptide solutions with {PC-SAFT}},}\ }\href@noop
  {} {\bibfield  {journal} {\bibinfo  {journal} {Chem. Eng. Process.}\ }\textbf
  {\bibinfo {volume} {47}},\ \bibinfo {pages} {1018} (\bibinfo {year}
  {2008})}\BibitemShut {NoStop}%
\bibitem [{\citenamefont {Wagner}\ and\ \citenamefont
  {Pru{\ss}}(2002)}]{Wagner2002}%
  \BibitemOpen
  \bibfield  {author} {\bibinfo {author} {\bibfnamefont {W.}~\bibnamefont
  {Wagner}}\ and\ \bibinfo {author} {\bibfnamefont {A.}~\bibnamefont
  {Pru{\ss}}},\ }\bibfield  {title} {\enquote {\bibinfo {title} {The {IAPWS}
  formulation 1995 for the thermodynamic properties of ordinary water substance
  for general and scientific use},}\ }\href@noop {} {\bibfield  {journal}
  {\bibinfo  {journal} {J. Phys. Chem. Ref. Data}\ }\textbf {\bibinfo {volume}
  {31}},\ \bibinfo {pages} {387} (\bibinfo {year} {2002})}\BibitemShut
  {NoStop}%
\bibitem [{\citenamefont {Tsonopoulos}(1999)}]{Tsonopoulos1999}%
  \BibitemOpen
  \bibfield  {author} {\bibinfo {author} {\bibfnamefont {C.}~\bibnamefont
  {Tsonopoulos}},\ }\bibfield  {title} {\enquote {\bibinfo {title}
  {Thermodynamic analysis of the mutual solubilities of normal alkanes and
  water},}\ }\href@noop {} {\bibfield  {journal} {\bibinfo  {journal} {Fluid
  Phase Equilib.}\ }\textbf {\bibinfo {volume} {156}},\ \bibinfo {pages} {21}
  (\bibinfo {year} {1999})}\BibitemShut {NoStop}%
\bibitem [{\citenamefont {Bell}, \citenamefont {Deiters},\ and\ \citenamefont
  {Leal}(2022)}]{Bell2022}%
  \BibitemOpen
  \bibfield  {author} {\bibinfo {author} {\bibfnamefont {I.~H.}\ \bibnamefont
  {Bell}}, \bibinfo {author} {\bibfnamefont {U.~K.}\ \bibnamefont {Deiters}},\
  and\ \bibinfo {author} {\bibfnamefont {A.~M.~M.}\ \bibnamefont {Leal}},\
  }\bibfield  {title} {\enquote {\bibinfo {title} {Implementing an equation of
  state without derivatives: teqp},}\ }\href@noop {} {\bibfield  {journal}
  {\bibinfo  {journal} {Ind. Eng. Chem. Res.}\ }\textbf {\bibinfo {volume}
  {61}},\ \bibinfo {pages} {6010} (\bibinfo {year} {2022})}\BibitemShut
  {NoStop}%
\bibitem [{\citenamefont {Rehner}, \citenamefont {Bauer},\ and\ \citenamefont
  {Gross}(2023)}]{Rehner2023}%
  \BibitemOpen
  \bibfield  {author} {\bibinfo {author} {\bibfnamefont {P.}~\bibnamefont
  {Rehner}}, \bibinfo {author} {\bibfnamefont {G.}~\bibnamefont {Bauer}},\ and\
  \bibinfo {author} {\bibfnamefont {J.}~\bibnamefont {Gross}},\ }\bibfield
  {title} {\enquote {\bibinfo {title} {{FeOs}: An open-source framework for
  equations of state and classical density functional theory},}\ }\href@noop {}
  {\bibfield  {journal} {\bibinfo  {journal} {Ind. Eng. Chem. Res.}\ }\textbf
  {\bibinfo {volume} {62}},\ \bibinfo {pages} {5347} (\bibinfo {year}
  {2023})}\BibitemShut {NoStop}%
\bibitem [{\citenamefont {Mulero}, \citenamefont {{Cachadi\~na}},\ and\
  \citenamefont {Parra}(2012)}]{Mulero2012}%
  \BibitemOpen
  \bibfield  {author} {\bibinfo {author} {\bibfnamefont {A.}~\bibnamefont
  {Mulero}}, \bibinfo {author} {\bibfnamefont {I.}~\bibnamefont
  {{Cachadi\~na}}},\ and\ \bibinfo {author} {\bibfnamefont {M.~I.}\
  \bibnamefont {Parra}},\ }\bibfield  {title} {\enquote {\bibinfo {title}
  {Recommended correlations for the surface tension of common fluids},}\
  }\href@noop {} {\bibfield  {journal} {\bibinfo  {journal} {J. Phys. Chem.
  Ref. Data}\ }\textbf {\bibinfo {volume} {41}},\ \bibinfo {pages} {043105}
  (\bibinfo {year} {2012})}\BibitemShut {NoStop}%
\bibitem [{\citenamefont {Goebel}\ and\ \citenamefont
  {Lunkenheimer}(1997)}]{Goebel1997}%
  \BibitemOpen
  \bibfield  {author} {\bibinfo {author} {\bibfnamefont {A.}~\bibnamefont
  {Goebel}}\ and\ \bibinfo {author} {\bibfnamefont {K.}~\bibnamefont
  {Lunkenheimer}},\ }\bibfield  {title} {\enquote {\bibinfo {title}
  {Interfacial tension of the water/n-alkane interface},}\ }\href@noop {}
  {\bibfield  {journal} {\bibinfo  {journal} {Langmuir}\ }\textbf {\bibinfo
  {volume} {13}},\ \bibinfo {pages} {369} (\bibinfo {year} {1997})}\BibitemShut
  {NoStop}%
\bibitem [{\citenamefont {Rogers}, \citenamefont {Leong},\ and\ \citenamefont
  {Wang}(2016)}]{Rogers2016}%
  \BibitemOpen
  \bibfield  {author} {\bibinfo {author} {\bibfnamefont {T.~R.}\ \bibnamefont
  {Rogers}}, \bibinfo {author} {\bibfnamefont {K.-Y.}\ \bibnamefont {Leong}},\
  and\ \bibinfo {author} {\bibfnamefont {F.}~\bibnamefont {Wang}},\ }\bibfield
  {title} {\enquote {\bibinfo {title} {Possible evidence for a new form of
  liquid buried in the surface tension of supercooled water},}\ }\href@noop {}
  {\bibfield  {journal} {\bibinfo  {journal} {Sci. Rep.}\ }\textbf {\bibinfo
  {volume} {6}},\ \bibinfo {pages} {33284} (\bibinfo {year}
  {2016})}\BibitemShut {NoStop}%
\bibitem [{\citenamefont {Tian}\ \emph {et~al.}(2014)\citenamefont {Tian},
  \citenamefont {Cao}, \citenamefont {Qiu},\ and\ \citenamefont
  {Zhu}}]{Tian2014}%
  \BibitemOpen
  \bibfield  {author} {\bibinfo {author} {\bibfnamefont {Y.}~\bibnamefont
  {Tian}}, \bibinfo {author} {\bibfnamefont {L.}~\bibnamefont {Cao}}, \bibinfo
  {author} {\bibfnamefont {L.}~\bibnamefont {Qiu}},\ and\ \bibinfo {author}
  {\bibfnamefont {R.}~\bibnamefont {Zhu}},\ }\bibfield  {title} {\enquote
  {\bibinfo {title} {A comparison study of the temperature dependence of
  interfacial tension in \textit{n}-alkane/water and \textit{n}-alcohol/water
  binary systems},}\ }\href@noop {} {\bibfield  {journal} {\bibinfo  {journal}
  {J. Chem. Eng. Data}\ }\textbf {\bibinfo {volume} {59}},\ \bibinfo {pages}
  {3495} (\bibinfo {year} {2014})}\BibitemShut {NoStop}%
\bibitem [{\citenamefont {Zeppieri}, \citenamefont {Rodr\'iguez},\ and\
  \citenamefont {{L\'opez de Ramos}}(2001)}]{Zeppieri2001}%
  \BibitemOpen
  \bibfield  {author} {\bibinfo {author} {\bibfnamefont {S.}~\bibnamefont
  {Zeppieri}}, \bibinfo {author} {\bibfnamefont {J.}~\bibnamefont
  {Rodr\'iguez}},\ and\ \bibinfo {author} {\bibfnamefont {A.~L.}\ \bibnamefont
  {{L\'opez de Ramos}}},\ }\bibfield  {title} {\enquote {\bibinfo {title}
  {Interfacial tension of alkane + water systems},}\ }\href@noop {} {\bibfield
  {journal} {\bibinfo  {journal} {J. Chem. Eng. Data}\ }\textbf {\bibinfo
  {volume} {46}},\ \bibinfo {pages} {1086} (\bibinfo {year}
  {2001})}\BibitemShut {NoStop}%
\bibitem [{\citenamefont {Girifalco}\ and\ \citenamefont
  {Good}(1957)}]{Girifalco1957}%
  \BibitemOpen
  \bibfield  {author} {\bibinfo {author} {\bibfnamefont {L.~A.}\ \bibnamefont
  {Girifalco}}\ and\ \bibinfo {author} {\bibfnamefont {R.~J.}\ \bibnamefont
  {Good}},\ }\bibfield  {title} {\enquote {\bibinfo {title} {A theory for the
  estimation of surface and interfacial energies. {I}. {D}erivation and
  application to interfacial tension},}\ }\href@noop {} {\bibfield  {journal}
  {\bibinfo  {journal} {J. Phys. Chem.}\ }\textbf {\bibinfo {volume} {61}},\
  \bibinfo {pages} {904} (\bibinfo {year} {1957})}\BibitemShut {NoStop}%
\bibitem [{\citenamefont {Fowkes}(1964)}]{Fowkes1964}%
  \BibitemOpen
  \bibfield  {author} {\bibinfo {author} {\bibfnamefont {F.~M.}\ \bibnamefont
  {Fowkes}},\ }\bibfield  {title} {\enquote {\bibinfo {title} {Attractive
  forces at interfaces},}\ }\href@noop {} {\bibfield  {journal} {\bibinfo
  {journal} {Ind. Eng. Chem.}\ }\textbf {\bibinfo {volume} {56}},\ \bibinfo
  {pages} {40} (\bibinfo {year} {1964})}\BibitemShut {NoStop}%
\bibitem [{\citenamefont {Bonn}\ and\ \citenamefont
  {Ross}(2001)}]{BonnRoss2001}%
  \BibitemOpen
  \bibfield  {author} {\bibinfo {author} {\bibfnamefont {D.}~\bibnamefont
  {Bonn}}\ and\ \bibinfo {author} {\bibfnamefont {D.}~\bibnamefont {Ross}},\
  }\bibfield  {title} {\enquote {\bibinfo {title} {Wetting transitions},}\
  }\href@noop {} {\bibfield  {journal} {\bibinfo  {journal} {Rep. Prog. Phys.}\
  }\textbf {\bibinfo {volume} {64}},\ \bibinfo {pages} {1085} (\bibinfo {year}
  {2001})}\BibitemShut {NoStop}%
\bibitem [{\citenamefont {Caupin}\ and\ \citenamefont
  {Anisimov}(2019)}]{Caupin2019}%
  \BibitemOpen
  \bibfield  {author} {\bibinfo {author} {\bibfnamefont {F.}~\bibnamefont
  {Caupin}}\ and\ \bibinfo {author} {\bibfnamefont {M.~A.}\ \bibnamefont
  {Anisimov}},\ }\bibfield  {title} {\enquote {\bibinfo {title} {Thermodynamics
  of supercooled and stretched water: Unifying two-structure description and
  liquid-vapor spinodal},}\ }\href@noop {} {\bibfield  {journal} {\bibinfo
  {journal} {J. Chem. Phys.}\ }\textbf {\bibinfo {volume} {151}},\ \bibinfo
  {pages} {034503} (\bibinfo {year} {2019})}\BibitemShut {NoStop}%
\bibitem [{\citenamefont {Novak}, \citenamefont {Liang},\ and\ \citenamefont
  {Kontogeorgis}(2025)}]{Novak2025}%
  \BibitemOpen
  \bibfield  {author} {\bibinfo {author} {\bibfnamefont {N.}~\bibnamefont
  {Novak}}, \bibinfo {author} {\bibfnamefont {X.}~\bibnamefont {Liang}},\ and\
  \bibinfo {author} {\bibfnamefont {G.~M.}\ \bibnamefont {Kontogeorgis}},\
  }\bibfield  {title} {\enquote {\bibinfo {title} {{PC-SAFT-TS-CAF}: a revised
  two state model for pure water and binary aqueous mixtures},}\ }\href@noop {}
  {\bibfield  {journal} {\bibinfo  {journal} {Fluid Phase Equilib.}\ }\textbf
  {\bibinfo {volume} {598}},\ \bibinfo {pages} {114475} (\bibinfo {year}
  {2025})}\BibitemShut {NoStop}%
\end{thebibliography}%

\end{document}